\documentclass[a4paper,12pt]{article}

\pdfoutput=1
\usepackage{amsmath}
\usepackage{amssymb}
\usepackage{amsfonts}
\usepackage{mathrsfs}
\usepackage{bbm}
\usepackage{graphicx,subfigure}

\usepackage{bookmark}

\usepackage{cite}

\usepackage{calc}
\usepackage{hyperref}
\usepackage{array}

\usepackage{ulem}

\usepackage{multirow}
\usepackage{color}
\usepackage{xcolor,colortbl}

\usepackage{psfrag}
\usepackage{pstricks}
\usepackage{epsfig}

\allowdisplaybreaks

\newlength{\dinwidth}
\newlength{\dinmargin}
\definecolor{nicered}{rgb}{1.0,0.0,0.2}

\definecolor{color1}{rgb}{0.9,.4,.2}

\definecolor{color2}{rgb}{0.3,.6,.7}

\definecolor{color3}{rgb}{0.7,.2,.7}

\usepackage{color}

\begin{document}

\title{
\vspace*{-0.5cm}
\bf \Large
Estimating of CP Violation in $B_{c}\rightarrow B K^{0}+B {\bar{K}}^0\rightarrow B \pi^{\pm} e^{\mp} \nu_{e}$ Decays}

\author{Xiao-Dong Cheng$^{1}$\footnote{chengxd@mails.ccnu.edu.cn}, Zhen-Lu Weng$^{1}$\footnote{252844791@qq.com}, Ying-Ying Fan$^{1}$\footnote{fyy163@126.com}, Ru-Min Wang$^{2}$\footnote{ruminwang@sina.com}\\
\\
{$^1$\small College of Physics and Electronic Engineering,}\\[-0.2cm]
{    \small Xinyang Normal University, Xinyang 464000, People's Republic of China}\\[-0.1cm]
{$^2$\small College of Physics and Communication Electronics,}\\[-0.2cm]
{    \small JiangXi Normal University, NanChang 330022, People's Republic of China}\\[-0.1cm]}

\date{}
\maketitle
\bigskip\bigskip
\maketitle
\vspace{-1.2cm}

\begin{abstract}
{\noindent}In this paper, we investigate the CP asymmetries ${\mathcal A}_{CP}^{pm}$ and ${\mathcal A}_{CP}^{mp}$ in $B_{c}^{\pm}\rightarrow B^{\pm} K^{0}+B^{\pm} {\bar{K}}^0\rightarrow  B^{\pm} \pi^{\pm} e^{\mp} \nu_{e}$ and $B_{c}^{\pm}\rightarrow B^{\pm} K^{0}+B^{\pm} {\bar{K}}^0\rightarrow  B^{\pm} \pi^{\mp} e^{\pm} \nu_{e}$ decays, both of them consist of three parts: the indirect CP violations in $K^0 -\bar{K}^0$ mixing $A_{CP,mix}^{pm}$ and $A_{CP,mix}^{mp}$, the direct CP violations in $B_{c}$ decay $A_{CP,dir}^{pm}$ and $A_{CP,dir}^{mp}$, the new CP violation effects $A_{CP,int}^{pm}$ and $A_{CP,int}^{mp}$, which originate from the interference between the amplitude of the $B_{c}^{-}\rightarrow B^{-} K^{0} ({\bar{K}}^0) \rightarrow B^{-} \pi^{-}  e^{+} \nu_{e}(\pi^{+}  e^{-} \bar{\nu}_{e})$ decay with the difference between the oscillating effect of $K^0\rightarrow {\bar{K}}^0$ and that of ${\bar{K}}^0\rightarrow K^0 $. The strong phase differences of the direct CP asymmetries $A_{CP,dir}^{pm}$ and $A_{CP,dir}^{mp}$ arising from $K^0-\bar{K}^0$ mixing parameters. ${\mathcal A}_{CP}^{pm}$ and ${\mathcal A}_{CP}^{mp}$ are of the order $\mathcal{O}(10^{-4})$ and $\mathcal{O}(10^{-3})$, respectively. The new CP violation effect ${\mathcal A}_{CP,int}^{pm}$ plays a dominant role in ${\mathcal A}_{CP}^{pm}$, the CP asymmetry ${\mathcal A}_{CP}^{mp}$ is dominated by the indirect CP violation ${\mathcal A}_{CP,mix}^{mp}$, so the CP asymmetry ${\mathcal A}_{CP}^{pm}$ provides an ideal place to study the new CP violation effect. We derive another two asymmetry observables ${\mathcal A}_{CP}^{pp}$ and ${\mathcal A}_{CP}^{mm}$, which are dominated by $K^0-\bar{K}^0$ mixing. The observables ${\mathcal A}_{CP}^{pm}$, ${\mathcal A}_{CP}^{mp}$, ${\mathcal A}_{CP}^{pp}$ and ${\mathcal A}_{CP}^{mm}$ are hopefully to be marginally observed at the LHC experiment and the HL-LHC experiment.
\end{abstract}
\newpage

\section{Introduction}
\label{sec:intro}
In the Standard Model of particle physics, CP violation originates from a nonzero phase parameter within the Cabibbo-Kobayashi-Mashawa (CKM) matrix~\cite{Cabibbo:1963yz,Kobayashi:1973fv}. In general, the CKM mechanism is very successful in describing all observed CP violation phenomena and is, very likely, the dominant source of CP violation in low energy flavor-changing processes~\cite{HFLAV:2022esi,Nir:2005js}. However, the CKM mechanism fails to account for the astronomical observations suggested amount of matter-antimatter asymmetry, which presents a significant challenge to the Standard Model and hints at the presence of additional sources of CP violation~\cite{Dine:2003ax}. Probing CP violation in various reactions and seeing the correlation between different processes may open a window to the discovery of additional sources of CP violation~\cite{Li:2005ci}.

CP violation has been well established in both $K$ and $B$ systems, however, the only place where CP violation was discovered in charmed systems was the difference between the time-integrated asymmetries of $D^0\rightarrow K^+ K^-$ and $D^0\rightarrow {\pi}^+ {\pi}^-$ at $5.3\sigma$ level, which is reported by the LHCb collaboration in 2019~\cite{Aaij:2019kcg}. On the other hand, many theoretical efforts have so far been made to study CP violations in the D system and charm baryon during the past decade~\cite{Buccella:1992sg,Buccella:1994nf,Cheng:2025kpp,Cheng:2021yrn,Cheng:2012wr,Cheng:2012xb,Li:2021iwf,Zhang:2021zhr,Lenz:2020awd,Unal:2020ezc, Jia:2024pyb,Wang:2022tcm,Saur:2020rgd,Wang:2017gxe,Yu:2017oky,Qin:2013tje,Li:2012cfa,Song:2025lmj,Azimov:1999gw,Azimov:1999dj,Lipkin:1999qz,Xing:1995jg,Wang:2022hop}.
However, studies of CP-violating processes in the doubly heavy-flavored $B_c$ decays that proceed via the c quark decays with the b quark as a spectator are very scarce~\cite{Cheng:2021yfr}.

The decays with final states including $K^0$ or ${\bar{K}}^0$ can be used to study several distinctive CP-violation effects, such as the indirect CP violation induced by the $K^0-{\bar{K}}^0$ mixing~\cite{Amorim:1998pi,CDF:2012lpc,Thomas:2012qf,BaBar:2012wep,Belle:2012ygx,Wang:2017gxe,Yu:2017oky,Ko:2012pe,Ko:2010ng,delAmoSanchez:2011zza,BABAR:2011aa,Mendez:2009aa,Dobbs:2007ab,Link:2001zj,Grossman:2011zk,Bigi:2012km,
Chen:2021udz,Chen:2020uxi,Chen:2019vbr,Dighe:2019odu,Rendon:2019awg,Cirigliano:2019wxv,Castro:2018cot,Delepine:2018amd,Cirigliano:2017tqn,Dhargyal:2016kwp,Devi:2013gya,Kimura:2014wsa}. Moreover, due to  the $\Delta S =\Delta Q$ rule, the semi-leptonic $K^0$ and ${\bar{K}}^0$ transitions may have rich physics and provide a clean environment to study CP violation. In this paper, we calculate the CP violations in $B_{c}^{\pm}\rightarrow B^{\pm} K^{0}+B^{\pm} {\bar{K}}^0\rightarrow B^{\pm} \pi^{\pm} e^{\mp} \nu_{e}$ and $B_{c}^{\pm}\rightarrow B^{\pm} K^{0}+B^{\pm} {\bar{K}}^0\rightarrow B^{\pm} \pi^{\mp} e^{\pm} \nu_{e}$ decays. We predict the branching ratios of these decays and investigate the CP asymmetries, which have three kinds of CP violation effects. The paper is organized as follows. In section~\ref{sec:decaywidth}, we derive the branching ratios of the $B_{c}^{\pm}\rightarrow B^{\pm} K^{0}+B^{\pm} {\bar{K}}^0\rightarrow B^{\pm} \pi^{\pm} e^{\mp} \nu_{e}$ and $B_{c}^{\pm}\rightarrow B^{\pm} K^{0}+B^{\pm} {\bar{K}}^0\rightarrow B^{\pm} \pi^{\mp} e^{\pm} \nu_{e}$ decays. In section~\ref{sec:cpviolationcal}, we calculate the CP asymmetry observables for the $B_{c}^{\pm}\rightarrow B^{\pm} K^{0}+B^{\pm} {\bar{K}}^0\rightarrow B^{\pm} \pi^{\pm} e^{\mp} \nu_{e}$ and $B_{c}^{\pm}\rightarrow B^{\pm} K^{0}+B^{\pm} {\bar{K}}^0\rightarrow B^{\pm} \pi^{\mp} e^{\pm} \nu_{e}$ decays. The numerical results and discussions are presented in section~\ref{sec:numberres}, we also investigate the sensitivity of these measurements in this section. And section~\ref{sec:conclusions} is the conclusion.
\section{Branching fractions of the $B_{c}\rightarrow B K^{0}+B {\bar{K}}^0\rightarrow B \pi^{\pm} e^{\mp} \nu_{e}$ decays}
\label{sec:decaywidth}
\begin{figure}[t]
\centering
\hspace{0cm}\includegraphics[width=0.36\textwidth]{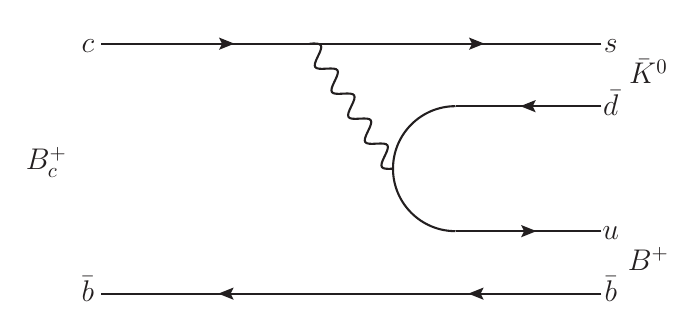}
\hspace{2cm}\includegraphics[width=0.36\textwidth]{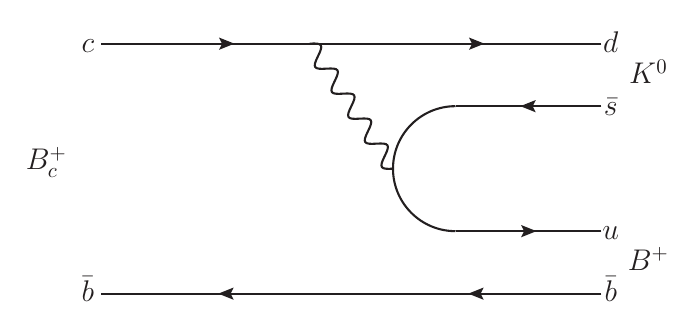}
\caption{\small Quark diagrams for the $B_c^+\rightarrow B^{+} \bar{K}^{0}$ and $B_c^+\rightarrow B^{+} K^{0}$ decays.}
\label{cabibbofavorbcbukz}
\end{figure}

\begin{figure}[t]
\centering
\hspace{0cm}\includegraphics[width=0.36\textwidth]{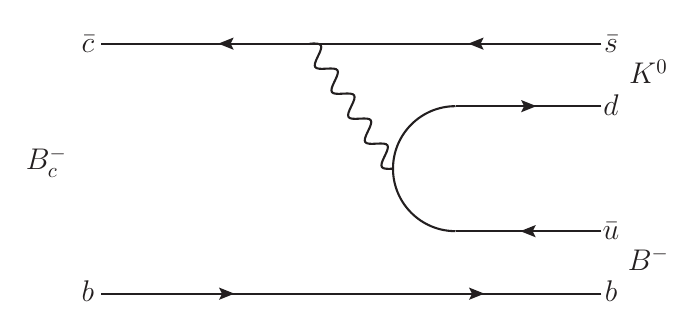}
\hspace{2cm}\includegraphics[width=0.36\textwidth]{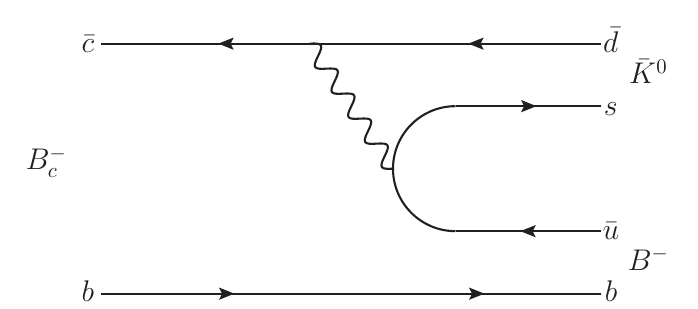}
\caption{\small Quark diagrams for the $B_c^-\rightarrow B^{-} K^{0}$ and $B_c^-\rightarrow B^{-} \bar{K}^{0}$ decays.}
\label{dousupcabibbcbukzb}
\end{figure}

Within the Standard Model, the $B_c^+\rightarrow B^{+} \bar{K}^{0}$ and $B_c^+\rightarrow B^{+} K^{0}$ decays are generated by $c\rightarrow s u \bar{d}$ and $c\rightarrow d u \bar{s}$ transitions, respectively. Their charge conjugate decays $B_c^-\rightarrow B^{-} K^{0}$ and $B_c^-\rightarrow B^{-} \bar{K}^{0}$ can occur through $\bar{c}\rightarrow \bar{s} \bar{u} d$ and $\bar{c}\rightarrow \bar{d} \bar{u} s$ transitions, respectively. The Feynman diagrams of these decays are shown in Fig~\ref{cabibbofavorbcbukz} and Fig~\ref{dousupcabibbcbukzb}. The  effective Hamiltonian for the $B_c^{\pm}\rightarrow B^{\pm} K^0 (\bar{K}^{0})$ decays is
\begin{align}
\mathcal H_{\rm eff}=\frac{G_F}{\sqrt{2}} \left[V_{cs}^{\ast}V_{ud} \bar{s}\gamma^{\mu}(1-\gamma_{5})c\hspace{0.05cm}\bar{u}\gamma_{\mu}(1-\gamma_{5})d+V_{cd}^{\ast}V_{us} \bar{d}\gamma^{\mu}(1-\gamma_{5})c\hspace{0.05cm}\bar{u}\gamma_{\mu}(1-\gamma_{5})s \right]+\text{h.c.}, \label{Eq:effhamilton1}
\end{align}
where $G_F$ is the Fermi coupling constant, $V_{ij}$ denotes the CKM matrix element. To calculate the hadronic matrix elements of the effective Hamiltonian, we need to reconstruct the effective Hamiltonian in a form suitable for the use of the factorization assumption by Fierz transformation~\cite{Cheng:2012wr,Fakirov:1977ta,Wang:2008tf,Giri:1998qf}. By making use of the Fierz transformation, we obtain the factorized Hamiltonian
\begin{align}
\mathcal H_{\rm eff}^{fac}=-\frac{G_F}{\sqrt{2} N_{c}} \left[V_{cs}^{\ast}V_{ud}\bar{s}\gamma^{\mu}(1-\gamma_{5})d\hspace{0.05cm}\bar{u}\gamma_{\mu}(1-\gamma_{5})c +V_{cd}^{\ast}V_{us}\bar{d}\gamma^{\mu}(1-\gamma_{5})s\hspace{0.05cm}\bar{u}\gamma_{\mu}(1-\gamma_{5})c \right]+\text{h.c.}, \label{Eq:effhamilton2}
\end{align}
with $N_c=3$ is the number of colors. According to the factorized Hamiltonian, the hadronic matrix elements $\left \langle K^0 (q_t)B^{\pm}(p_2) \left|\mathcal H_{\rm eff}^{fac}\right|B^{\pm}_{c}(p_1)\right\rangle$ and $\left \langle \bar{K}^0 (q_t)B^{\pm}(p_2) \left|\mathcal H_{\rm eff}^{fac}\right|B^{\pm}_{c}(p_1)\right\rangle$ can be written as~\cite{Cheng:2021yfr}
\begin{align}
\left \langle K^0 (q_t)B^{+}(p_2) \left|\mathcal H_{\rm eff}^{fac}\right|B^{+}_{c}(p_1)\right\rangle=\frac{{G_F} f_K}{3\sqrt{2}}V_{cd}^{\ast}V_{us} e^{i\phi} f_0(q_t^2) \left(p_{1}^2-p_{2}^2\right),\label{Eq:hadmatrixele1}\\
\left \langle \bar{K}^0 (q_t)B^{+}(p_2) \left|\mathcal H_{\rm eff}^{fac}\right|B^{+}_{c}(p_1)\right\rangle=\frac{{G_F} f_K}{3\sqrt{2}}V_{cs}^{\ast}V_{ud} e^{i\phi} f_0(q_t^2) \left(p_{1}^2-p_{2}^2\right),\label{Eq:hadmatrixele2}\\
\left \langle K^0 (q_t)B^{-}(p_2) \left|\mathcal H_{\rm eff}^{fac}\right|B^{-}_{c}(p_1)\right\rangle=\frac{{G_F} f_K}{3\sqrt{2}}V_{cs}V_{ud}^{\ast} e^{i\phi} f_0(q_t^2) \left(p_{1}^2-p_{2}^2\right),\label{Eq:hadmatrixele3}\\
\left \langle \bar{K}^0 (q_t)B^{-}(p_2) \left|\mathcal H_{\rm eff}^{fac}\right|B^{-}_{c}(p_1)\right\rangle=\frac{{G_F} f_K}{3\sqrt{2}}V_{cd}V_{us}^{\ast} e^{i\phi} f_0(q_t^2) \left(p_{1}^2-p_{2}^2\right),\label{Eq:hadmatrixele4}
\end{align}
where $\phi$ denotes the strong phase of the involved matrix elements, which are identical in the $B^{\pm}_{c}\rightarrow B^{\pm} K^0  $ and $B^{\pm}_{c}\rightarrow B^{\pm} \bar{K}^0$ decays. The reason is as follows: both the $B^{\pm}_{c}\rightarrow B^{\pm} K^0  $ and $B^{\pm}_{c}\rightarrow B^{\pm} \bar{K}^0$ decays are simple, they receive only the contribution of the color-suppressed tree-emission diagram, without the contributions of the color-allowed tree-emission diagram, the W-annihilation diagram and the W-exchange diagram. In this case, the strong phases in the Cabibbo-favored decay (the $B^{+}_{c}\rightarrow B^{+} \bar{K}^0$ and $B^{-}_{c}\rightarrow B^{-} K^0  $ decays) and the doubly Cabibbo-suppressed decay (the $B^{-}_{c}\rightarrow B^{-} \bar{K}^0$ and $B^{+}_{c}\rightarrow B^{+} K^0  $ decays) are identical~\cite{Li:2012cfa,Cheng:2021yfr}, so the presence of the strong phase $\phi$ in the involved matrix elements will not change the results of the branching ratios and the CP asymmetries of the $B^{\pm}_{c}\rightarrow B^{\pm} K^0  $ and $B^{\pm}_{c}\rightarrow B^{\pm} \bar{K}^0$ decays, which are related to the products of the matrix elements in Eqs.(\ref{Eq:hadmatrixele1})-(\ref{Eq:hadmatrixele4}) and the complex conjugates of them. $f_K$ is the decay constant of $K^0$ and is defined by~\cite{Branco:1999fs,ExtendedTwistedMass:2020tvp}
\begin{align}
\left \langle K^0 (q_t) \left|\bar{d} \gamma^\mu \gamma_5 s\right|0\right\rangle=iq_t^\mu f_{K}, \label{Eq:kdecaycons}\\
\left \langle |\bar{K}^0 (q_t) \left|\bar{s} \gamma^\mu \gamma_5 d\right|0\right\rangle=iq_t^\mu f_{K}, \label{Eq:kbdecaycons}
\end{align}
$q_t=p_1-p_2$ is the 4-momentum transfer, $f_0(q_t^2)$ is one of two form factors that parameterize the matrix elements~\cite{Cooper:2020wnj}
\begin{align}
\left \langle B^{+} (p_2) \left|\bar{u} \gamma^\mu  c\right|B^+_{c}(p_1)\right\rangle=f_0(q_t^2)\left(\frac{p_{1}^2-p_{2}^2}{q_t^2}{q_t^{\mu}}\right)+f_{+}(q_t^2)\left(p_1^{\mu}+p_2^{\mu}-\frac{p_{1}^2-p_{2}^2}{q_t^2}{q_t^{\mu}}\right),\label{Eq:formfactor1} \\
\left \langle B^{-} (p_2) \left|\bar{c} \gamma^\mu  u\right|B^-_{c}(p_1)\right\rangle=-f_0(q_t^2)\left(\frac{p_{1}^2-p_{2}^2}{q_t^2}{q_t^{\mu}}\right)-f_{+}(q_t^2)\left(p_1^{\mu}+p_2^{\mu}-\frac{p_{1}^2-p_{2}^2}{q_t^2}{q_t^{\mu}}\right),  \label{Eq:formfactor2}
\end{align}
here, the form factors can be represented as
\begin{align}
f_0(q_t^2)=\frac{1}{1-\frac{q_t^2}{m_{D_0^\ast}^2}}\left(b_0^{(0)}+b_0^{(1)}Z_p+b_0^{(2)}Z_p^2+b_0^{(3)}Z_p^3\right), \label{Eq:fzero}\\
f_+(q_t^2)=\frac{1}{1-\frac{q_t^2}{m_{D^\ast}^2}}\left(b_+^{(0)}+b_+^{(1)}Z_p+b_+^{(2)}Z_p^2+b_+^{(3)}Z_p^3\right), \label{Eq:fplus}
\end{align}
$m_{D_0^\ast}$ and $m_{D^\ast}$ denote respectively the mass of the neutral scalar meson $D_0^\ast(2300)^0$ and the neutral vector  meson $D^{\ast}(2007)^0$. The variable $Z_p$ is defined as
\begin{align}
Z_p=\left(\frac{\sqrt{t_{+}-q_t^2}-\sqrt{t_+}}{\sqrt{t_{+}-q_t^2}+\sqrt{t_+}}\right)\cdot\left|\frac{\sqrt{t_{+}-m_{D^\ast}^2}+\sqrt{t_+}}{\sqrt{t_{+}-m_{D^\ast}^2}-\sqrt{t_+}}\right|, \label{Eq:zpfunction}
\end{align}
with $t_+={\left(m_{B_c^+}+m_{B^+}\right)}^2$. The polynomial coefficients $b_{0,+}^{(i)} (i=0,1,2,3)$ appearing in  Eq.(\ref{Eq:fzero}) and Eq.(\ref{Eq:fplus}) have been calculated in Ref.~\cite{Cooper:2020wnj}
\begin{align}
b_{0}^{(0)}=0.548\pm0.023,~~~~~b_{0}^{(1)}=-0.19\pm0.22,~~~~~b_{0}^{(2)}=0.05\pm0.74,~~~~~ b_{0}^{(3)}=0, \label{Eq:coefficients1}\\
b_{+}^{(0)}=0.548\pm0.023,~~~~~b_{+}^{(1)}=-0.48\pm0.21,~~~~~b_{+}^{(2)}=0.12\pm0.77,~~~~~ b_{+}^{(3)}=0.\label{Eq:coefficients2}
\end{align}
Under the assumption that CPT is invariant, the time-evolved state of an initially $(t=0)$ pure  $K^0$ or $\bar{K}^0$ is given as~\cite{ParticleDataGroup:2024cfk}
\begin{align}
&\left | K^0_{phys} (t) \right\rangle=g_{+}(t) \left | K^0 \right\rangle - \frac{q}{p} g_{-}(t) \left | \bar{K}^0 \right\rangle,\label{Eq:kzphysdef}\\
&\left | \bar{K}^0_{phys} (t) \right\rangle=-\frac{p}{q} g_{-}(t) \left | K^0\right\rangle + g_{+} (t) \left | \bar{K}^0 \right\rangle.\label{Eq:kbphysdef}
\end{align}
with
\begin{align}
g_{\pm}(t)=\frac{1}{2} (e^{-i m_L t -\frac{1}{2}\Gamma_L t }\pm e^{-i m_S t -\frac{1}{2}\Gamma_S t }),\label{Eq:gpmtcoeff}
\end{align}
where $m_L$ and $m_S$ are the masses of the mass eigenstates $K^0_L$ and $K^0_S$, respectively. $\Gamma_L$ and $\Gamma_S$ are the widths of the mass eigenstates $K^0_L$ and $K^0_S$, respectively. The difference in masses of $K_L^0$ and $K_S^0$ and the average in widths of $K_L^0$ and $K_S^0$ are given by
\begin{align}
\Delta m=m_L-m_S,~~~~~~~~~~~~~~~~~~~~\Gamma=\frac{\Gamma_L +\Gamma_S}{2}.
\label{Eq:dimassavegamma}
\end{align}
$p$ and $q$ in Eq.(\ref{Eq:kzphysdef}) and Eq.(\ref{Eq:kbphysdef}) are complex mixing parameters and can be parameterized as
\begin{align}
p=\frac{1+\epsilon}{\sqrt{2(1+|\epsilon|^2)}},~~~~~~~~~~q=\frac{1-\epsilon}{\sqrt{2(1+|\epsilon|^2)}},\label{Eq:ksdefinition3}
\end{align}
where the complex parameter $\epsilon$ signifies deviation of the mass eigenstates from the CP eigenstates. Under the assumption that the $\Delta S =\Delta Q$ rule is valid, the amplitudes for the semileptonic decays $K^0\rightarrow \pi^{+} e^{-} \bar{\nu}_{e}$ and $\bar{K}^0 \rightarrow \pi^{-}  e^{+} \nu_{e}$ are given by
\begin{align}
&A\left(K^0\rightarrow \pi^{+} e^{-} \bar{\nu}_{e}\right)=0,\label{Eq:kzpipemnu}\\
&A\left(\bar{K}^0\rightarrow \pi^{-}  e^{+} \nu_{e}\right)=0.\label{Eq:kzbpimepnu}
\end{align}
By combining Eq.(\ref{Eq:kzphysdef}), Eq.(\ref{Eq:kbphysdef}), Eq.(\ref{Eq:kzpipemnu}) and  Eq.(\ref{Eq:kzbpimepnu}), we obtain the amplitudes for the  $K^0_{phys} (t)\rightarrow \pi^{\pm} e^{\mp}\nu_{e}$ and $\bar{K}^0_{phys} (t)\rightarrow \pi^{\pm} e^{\mp}\nu_{e}$ decays as follows:
\begin{align}
&A\left(K^0_{phys} (t)\rightarrow \pi^{+} e^{-} \bar{\nu}_{e}\right)=-\frac{q}{p} g_{-}(t) A\left(\bar{K}^0\rightarrow \pi^{+} e^{-} \bar{\nu}_{e}\right),\label{Eq:ampkphyspienv1}\\
&A\left(\bar{K}^0_{phys} (t)\rightarrow \pi^{-}  e^{+} \nu_{e}\right)=-\frac{p}{q} g_{-}(t) A\left(K^0\rightarrow \pi^{-} e^{+} \bar{\nu}_{e}\right),\label{Eq:ampkphyspienv2}\\
&A\left(K^0_{phys} (t)\rightarrow \pi^{-}  e^{+} \nu_{e}\right)= g_{+}(t) A\left(K^0\rightarrow \pi^{-}  e^{+} \nu_{e}\right),\label{Eq:ampkphyspienv3}\\
&A\left(\bar{K}^0_{phys} (t)\rightarrow \pi^{+} e^{-} \bar{\nu}_{e}\right)= g_{+}(t) A\left(\bar{K}^0\rightarrow \pi^{+} e^{-} \bar{\nu}_{e}\right),\label{Eq:ampkphyspienv4}
\end{align}
where $t$ refers to duration time of $K^0(\bar{K}^0)$ in the rest frame of $K^0(\bar{K}^0)$, which relates to the $K^0(\bar{K}^0)$ flight length in the LHCb detector~\cite{Song:2025lmj}.

Now, we proceed to consider the time-dependent amplitude of the cascade decay processes $B_{c}^{\pm}\rightarrow B^{\pm} K^{0}+B^{\pm} {\bar{K}}^0\rightarrow B^{\pm} \pi^{+} e^{-} \bar{\nu}_{e}$ and $B_{c}^{\pm}\rightarrow B^{\pm} K^{0}+B^{\pm} {\bar{K}}^0\rightarrow B^{\pm} \pi^{-} e^{+} \nu_{e}$, which can be written as
\begin{align}
&A\left(B^{\pm}_{c}\rightarrow B^{\pm}K^0(t)\rightarrow B^{\pm} \pi^{+} e^{-} \bar{\nu}_{e}\right)\nonumber\\
&~~~~~~~~~=\left \langle K^0 (q_t)B^{\pm}(p_2) \left|\mathcal H_{\rm eff}\right|B^{\pm}_{c}(p_1)\right\rangle \cdot A(K^0_{phys} (t)\rightarrow \pi^{+} e^{-} \bar{\nu}_{e})\nonumber\\
&~~~~~~~~~~+ \left \langle \bar{K}^0 (q_t)B^{\pm}(p_2) \left|\mathcal H_{\rm eff}\right|B^{\pm}_{c}(p_1)\right\rangle \cdot A(\bar{K}^0_{phys} (t)\rightarrow \pi^{+} e^{-} \bar{\nu}_{e}),\label{Eq:amtibcpmpipmemp1}\\
&A\left(B^{\pm}_{c}\rightarrow B^{\pm}K^0(t)\rightarrow B^{\pm} \pi^{-}  e^{+} \nu_{e}\right)\nonumber\\
&~~~~~~~~~=\left \langle K^0 (q_t)B^{\pm}(p_2) \left|\mathcal H_{\rm eff}\right|B^{\pm}_{c}(p_1)\right\rangle \cdot A(K^0_{phys} (t)\rightarrow \pi^{-}  e^{+} \nu_{e})\nonumber\\
&~~~~~~~~~~+ \left \langle \bar{K}^0 (q_t)B^{\pm}(p_2) \left|\mathcal H_{\rm eff}\right|B^{\pm}_{c}(p_1)\right\rangle \cdot A(\bar{K}^0_{phys} (t)\rightarrow \pi^{-}  e^{+} \nu_{e}).\label{Eq:amtibcpmpipmemp2}
\end{align}
Substituting Eqs.(\ref{Eq:hadmatrixele1})-(\ref{Eq:hadmatrixele4}) and Eqs.(\ref{Eq:ampkphyspienv1})-(\ref{Eq:ampkphyspienv4}) into Eq.(\ref{Eq:amtibcpmpipmemp1}) and Eq.(\ref{Eq:amtibcpmpipmemp2}), we obtain
\begin{align}
&A\left(B^{+}_{c}\rightarrow B^{+}K^0(t)\rightarrow B^{+} \pi^{+} e^{-} \bar{\nu}_{e}\right)\nonumber\\
&~~~=\frac{{G_F} f_K}{3\sqrt{2}} e^{i\phi} f_0(q_t^2) \left(p_{1}^2-p_{2}^2\right) A\left(\bar{K}^0\rightarrow \pi^{+} e^{-} \bar{\nu}_{e}\right)\left[-V_{cd}^{\ast}V_{us}\cdot \frac{q}{p} g_{-}(t)+ V_{cs}^{\ast}V_{ud} \cdot g_{+}(t)\right],\label{Eq:ampjtbcpmpipmemp1}
\end{align}
\begin{align}
&A\left(B^{-}_{c}\rightarrow B^{-}K^0(t)\rightarrow B^{-} \pi^{-}  e^{+} \nu_{e}\right)=\frac{{G_F} f_K}{3\sqrt{2}} e^{i\phi} f_0(q_t^2)\left(p_{1}^2-p_{2}^2\right)\cdot\nonumber\\
&~~~~~~\cdot  A\left(K^0\rightarrow \pi^{-}  e^{+} \nu_{e}\right)\left[-V_{cd}V_{us}^{\ast} \cdot \frac{q}{p} g_{-}(t)+ V_{cs}V_{ud}^{\ast}  \cdot g_{+}(t)+V_{cd}V_{us}^{\ast} \cdot g_{-}(t) \left(\frac{q}{p}- \frac{p}{q}\right)\right]\nonumber\\
&~~~=\frac{{G_F} f_K}{3\sqrt{2}} e^{i\phi} f_0(q_t^2)\left(p_{1}^2-p_{2}^2\right) A\left(K^0\rightarrow \pi^{-}  e^{+} \nu_{e}\right)\left[-V_{cd}V_{us}^{\ast} \cdot \frac{p}{q} g_{-}(t)+ V_{cs}V_{ud}^{\ast}  \cdot g_{+}(t)\right],\label{Eq:ampjtbcpmpipmemp2}
\end{align}
\begin{align}
&A\left(B^{+}_{c}\rightarrow B^{+}K^0(t)\rightarrow B^{+} \pi^{-}  e^{+} \nu_{e}\right)\nonumber\\
&~~~=\frac{{G_F} f_K}{3\sqrt{2}} e^{i\phi} f_0(q_t^2) \left(p_{1}^2-p_{2}^2\right) A\left(K^0\rightarrow \pi^{-}  e^{+} \nu_{e}\right)\left[V_{cd}^{\ast}V_{us}\cdot g_{+}(t)-  V_{cs}^{\ast}V_{ud} \cdot \frac{p}{q} g_{-}(t)\right],\label{Eq:ampjtbcpmpipmemp3}
\end{align}
\begin{align}
&A\left(B^{-}_{c}\rightarrow B^{-}K^0(t)\rightarrow B^{-} \pi^{+} e^{-} \bar{\nu}_{e}\right)=\frac{{G_F} f_K}{3\sqrt{2}} e^{i\phi} f_0(q_t^2) \left(p_{1}^2-p_{2}^2\right) \cdot\nonumber\\
&~~~~~~\cdot A\left(\bar{K}^0\rightarrow \pi^{+} e^{-} \bar{\nu}_{e}\right)\left[V_{cd}V_{us}^{\ast} \cdot g_{+}(t) - V_{cs}V_{ud}^{\ast}  \cdot \frac{p}{q} g_{-}(t)+V_{cs}V_{ud}^{\ast}  \cdot g_{-}(t)\left( \frac{p}{q} -\frac{q}{p}\right)\right]\nonumber\\
&~~~=\frac{{G_F} f_K}{3\sqrt{2}} e^{i\phi} f_0(q_t^2) \left(p_{1}^2-p_{2}^2\right) A\left(\bar{K}^0\rightarrow \pi^{+} e^{-} \bar{\nu}_{e}\right)\left[V_{cd}V_{us}^{\ast} \cdot g_{+}(t) - V_{cs}V_{ud}^{\ast}  \cdot \frac{q}{p} g_{-}(t)\right].\label{Eq:ampjtbcpmpipmemp4}
\end{align}
Here, we note that the interferences between the first two terms inside the square bracket in the second line of Eqs.(\ref{Eq:ampjtbcpmpipmemp1})-(\ref{Eq:ampjtbcpmpipmemp4}) are related to the direct CP violation, the strong phase differences in $B_{c}^{\pm}\rightarrow B^{\pm} K^{0}+B^{\pm} {\bar{K}}^0\rightarrow B^{\pm} \pi^{\pm} e^{\mp} \nu_e (\pi^{\mp} e^{\pm} \nu_e)$ decays are provided by the difference between $g_{+}(t)$ and $-\frac{q}{p} g_{-}(t)( -\frac{p}{q} g_{-}(t))$, i.e. the strong phase differences in $B_{c}^{\pm}\rightarrow B^{\pm} K^{0}+B^{\pm} {\bar{K}}^0\rightarrow B^{\pm} \pi^{\pm} e^{\mp} \nu_e (\pi^{\mp} e^{\pm} \nu_e)$ decays arising from $K^0-\bar{K}^0$ mixing, and thus prevent pollution from strong dynamics~\cite{Shen:2023nuw}. Meanwhile, the interference between the second and third terms inside the square bracket in the second line of Eq.(\ref{Eq:ampjtbcpmpipmemp2}) and the interference between the first and third terms inside the square bracket in the second line of Eq.(\ref{Eq:ampjtbcpmpipmemp4}) are related to a new CP violation effect, which will discussed in the next section.

Making use of Eq.(\ref{Eq:ampjtbcpmpipmemp1}) and performing integration over phase space and over $t\in \left[0,+\infty\right)$, we can obtain the time-independent decay width for the $B_{c}^{+}\rightarrow B^{+} K^{0}+B^{+} {\bar{K}}^0\rightarrow B^{+} \pi^{+} e^{-} \bar{\nu}_{e}$ (hereinafter for brevity referred to as $B_{c}^{+}\rightarrow  B^{+} \pi^{+} e^{-} \bar{\nu}_{e}$) decay
\begin{align}
&\Gamma(B_{c}^{+}\rightarrow  B^{+} \pi^{+} e^{-} \bar{\nu}_{e})=\frac{G_F^2 f_K^2 \left(f_0(m_{K^0}^2)\right)^2}{288\pi m_{B_c^+}^3 }  f(m_{B_c^+},m_{B^+} ,m_{K^0})  (m_{B_c^+}^2-m_{B^+}^2)^2 \left|V_{cs}\right|^2 \left|V_{ud}\right|^2 \nonumber\\
&~\cdot\Gamma(\bar{K}^{0}\rightarrow \pi^{+} e^{-} \bar{\nu}_{e})  \int_{0}^{\infty}\left|g_{+}(t)\right|^2 dt \left[ r_{wf}^{2} r_{g}\left|\frac{q}{p}\right|^2 +r_{wf} r_{sf} e^{i(\delta+\phi)}\frac{q}{p} + r_{wf} r_{sf} e^{-i(\delta+\phi)}\frac{q^{\ast}}{p^{\ast}}+1\right].
\label{Eq:decaywidthbct1}
\end{align}
In the above equation, the following substitutions are made for convenience
\begin{align}
& r_{wf} e^{i\phi}= - \frac{V_{cd}^{\ast}V_{us}}{V_{cs}^{\ast}V_{ud}}, ~~~~~~~~r_{g}=\frac{\int_{0}^{\infty}\left|g_{-}(t)\right|^2 dt}{\int_{0}^{\infty}\left|g_{+}(t)\right|^2 dt}, ~~~~~~~~r_{sf} e^{i\delta}= \frac{\int_{0}^{\infty}g_{-}(t) \left(g_{+}(t)\right)^{\ast} dt}{\int_{0}^{\infty}\left|g_{+}(t)\right|^2 dt},
\label{Eq:definrwfrsfrg}
\end{align}
with
\begin{align}
&\int_{0}^{\infty}\left|g_{+}(t)\right|^2 dt=\frac{1}{4} \left[\frac{\Gamma_{L}+\Gamma_{S}}{\Gamma_{L} \Gamma_{S}}+ \frac{2\Gamma}{\Gamma^2+(\Delta m)^2}\right],\label{Eq:gpmtintgrate1} \\
&\int_{0}^{\infty}\left|g_{-}(t)\right|^2 dt=\frac{1}{4} \left[\frac{\Gamma_{L}+\Gamma_{S}}{\Gamma_{L} \Gamma_{S}}- \frac{2\Gamma}{\Gamma^2+(\Delta m)^2}\right],
\label{Eq:gpmtintgrate2}\\
&\int_{0}^{\infty}\left(g_{-}(t)\right)^{\ast} g_{+}(t) dt=\frac{1}{4} \left[\frac{\Gamma_{S}-\Gamma_{L}}{\Gamma_{L} \Gamma_{S}}+ \frac{2i\Delta m}{\Gamma^2+(\Delta m)^2}\right],\label{Eq:gpmtintgrate3}
\end{align}
where $r_{sf}$, $r_{wf}$ and $r_{g}$ are positive numbers, $\phi$ is the weak phase difference. $m_r$ with $r=B_c^+, B^+, K^0$ in Eq.(\ref{Eq:decaywidthbct1}) denotes the mass of the resonance, the function $f(x,y,z)$ in Eq.(\ref{Eq:decaywidthbct1}) is defined as
\begin{align}
f(x,y,z)=\sqrt{x^4+y^4+z^4-2x^2 y^2-2x^2 z^2-2y^2 z^2}.
\label{Eq:phasespacefunc}
\end{align}
In Eqs.(\ref{Eq:decaywidthbct1}), (\ref{Eq:gpmtintgrate1}), (\ref{Eq:gpmtintgrate2}) and (\ref{Eq:gpmtintgrate3}), the upper limit of the integration over the duration time $t$ of $K^0(\bar{K}^0)$ (hereinafter for brevity referred to as $t_{up}$) is infinity, which is based on the following considerations:
\begin{enumerate}
\item The $K^0(\bar{K}^0)$ flight length is measured in the laboratory frame, whereas the duration time $t$ of $K^0(\bar{K}^0)$ is measured in the rest frame of $K^0(\bar{K}^0)$, it would become larger during the transformation from the rest frame of $K^0(\bar{K}^0)$ to the lab frame, the $K^0(\bar{K}^0)$ flight length not only relates to the duration time $t$ of $K^0(\bar{K}^0)$, but also relates to the energy and momentum of $B_{c}^{\pm}$ mesons in the laboratory frame and that of $K^0(\bar{K}^0)$ in the rest frame of $B_{c}^{\pm}$ mesons.
\item The $K^0(\bar{K}^0) \rightarrow \pi^{\mp} e^{\pm} \nu_{e}$ decays could be measured at LHCb~\cite{AlvesJunior:2018ldo}. There are three particularly important features in these decays. Firstly, the neutrino cannot be detected by the LHCb detector, so a good candidate for the $K^0(\bar{K}^0) \rightarrow \pi^{\mp} e^{\pm} \nu_{e}$ decays must have the missing momentum. Secondly, there are a charged pion and a charged lepton ($e^{\pm}$) in the final states of these decays, so the $K^0(\bar{K}^0) \rightarrow \pi^{\mp} e^{\pm} \nu_{e}$ candidate can be selected by requiring a well-identified charged pion and a high quality electron, moreover, the tracks of the charged pion and the electron are constrained to originate from a common vertex in the event selection. Thirdly, the invariant mass distribution of $\pi^{\pm}$ and $e^{\mp}$ originating from the $K^0(\bar{K}^0)$ decays must has a specific range and line shape, which are determined by the MC simulation, so a good candidate for the $K^0(\bar{K}^0) \rightarrow \pi^{\mp} e^{\pm} \nu_{e}$ decays requires the invariant mass distribution of $\pi^{\pm}$ and $e^{\mp}$ must lie in the specific range and have the specific line shape. Meanwhile, there are still additional experimental features which have not been included above. By considering these experimental features, the search for the $K^0(\bar{K}^0) \rightarrow \pi^{\mp} e^{\pm} \nu_{e}$ decays can be well performed by the experimental physicists at LHCb.
\item Because the invariant mass and line shape requirements of the $\pi^{\pm} e^{\mp}$ system can reject the backgrounds from the random combinations of $\pi^{\pm}$ and $e^{\mp}$ that do not originate from the $K^0(\bar{K}^0)$ decays, the dominant backgrounds are from the $K^0(\bar{K}^0)$ decays involving $\pi^{\pm}$ and $e^{\mp}$ in the final states, such as the $K_{S,L}^{0}\rightarrow \pi^{\pm} \mu^{\mp} {\nu}_{\mu}\rightarrow \pi^{\pm} e^{\mp} {\nu}_{e} {\bar{\nu}}_{\mu} {\nu}_{\mu}$, $K_{S,L}^{0}\rightarrow \pi^{+} \pi^{-} e^{+} e^{-}$, $K_{L}^{0}\rightarrow \pi^{0} \pi^{\pm} e^{\mp} {\nu}_{e}$, $K_{L}^{0}\rightarrow \gamma \pi^{\pm} e^{\mp} {\nu}_{e}$ and $K_{L}^{0}\rightarrow \gamma\pi^{\pm} \mu^{\mp} {\nu}_{\mu}\rightarrow \gamma\pi^{\pm} e^{\mp} {\nu}_{e} {\bar{\nu}}_{\mu} {\nu}_{\mu}$ decays~\cite{ParticleDataGroup:2024cfk}. The branching  ratios of the $K_{S,L}^{0}\rightarrow \pi^{+} \pi^{-} e^{+} e^{-}$, $K_{L}^{0}\rightarrow \pi^{0} \pi^{\pm} e^{\mp} {\nu}_{e}$, $K_{L}^{0}\rightarrow \gamma \pi^{\pm} e^{\mp} {\nu}_{e}$ and $K_{L}^{0}\rightarrow \gamma\pi^{\pm} \mu^{\mp} {\nu}_{\mu}\rightarrow \gamma\pi^{\pm} e^{\mp} {\nu}_{e} {\bar{\nu}}_{\mu} {\nu}_{\mu}$ decays much less than that of the $K_{S,L}^{0}\rightarrow \pi^{\pm} e^{\mp} {\nu}_{e}$ decays, so these backgrounds can be safely neglected. The branching ratios of the $K_{S,L}^{0}\rightarrow \pi^{\pm} \mu^{\mp} {\nu}_{\mu}\rightarrow \pi^{\pm} e^{\mp} {\nu}_{e} {\bar{\nu}}_{\mu} {\nu}_{\mu}$ decays have the same order of magnitude as the signal processes $K_{S,L}^{0}\rightarrow \pi^{\pm} e^{\mp} {\nu}_{e}$, however, duo to the mean life of muon is of the order of $10^{-6}\text{ s}$, the probability of the decays $\mu^{\pm}\rightarrow  e^{\pm} {\nu}_{e} {\nu}_{\mu}$ occur in the vertex detector (VELO) and the Tracker Turicensis (TT) is small, however, the tracks of $e^{\mp}$ can be well detected and reconstructed only when the tracking information from the VELO (or from the TT) to the T stations is available in LHCb, so the requirement of a high quality electron can suppress these backgrounds. In addition, there exist three neutrinos in the final states of the background processes, while the signal processes only comprise a neutrino, the requirement of the invariant mass and line shape of the $\pi^{\pm} e^{\mp}$ system can also suppress these backgrounds. In a word, the $K^0(\bar{K}^0) \rightarrow \pi^{\mp} e^{\pm} \nu_{e}$ decays can be searched for at the low background levels.
\item In the LHCb experiment, the tracks of $\pi^{\pm}$ and $e^{\mp}$ originating from the $K^0(\bar{K}^0) \rightarrow \pi^{\mp} e^{\pm} \nu_{e}$ decays can be well detected and reconstructed when the decay positions of $K^0(\bar{K}^0)$ are located in the vertex detector (VELO) and the Tracker Turicensis (TT). The center of TT is at about $2.5 m$ from the interaction point, this indicates that the flight lengths of $K^0(\bar{K}^0)$ are less than $2.5 m$ if the decay positions of $K^0(\bar{K}^0)$ are located in VELO and TT~\cite{LHCbSiliconTrackerGroup:2018wnm}. When these experimental features are taken into account, $t_{up}$ can be set to a specific value, such as $2/\Gamma_S$.
\item However, the flight lengths of $K^0(\bar{K}^0)$ in the LHCb experiment doesn't match with the specific value of $t_{up}$ in theory, for example, even if the duration time $t$ of $K^0(\bar{K}^0)$ is small in an event, the flight length of $K^0(\bar{K}^0)$ in this event can be longer than the distance from the interaction point to the center of TT when the energy and momentum of $B_{c}^{\pm}$ mesons in the laboratory frame are large enough. On the other side, even if the flight length of $K^0(\bar{K}^0)$ in an event is less than $2.5 m$, the duration time $t$ of $K^0(\bar{K}^0)$ can be larger than the specific value of $t_{up}$ when the energy and momentum of $B_{c}^{\pm}$ mesons in the laboratory frame are small enough in this event. In a word, if a specific value of $t_{up}$ is adopted, the theoretical prediction, which is based on the specific value of $t_{up}$, can't match with the LHCb data sample directly.
\item There is a method to match the theoretical prediction with the LHCb data sample when a specific value of $t_{up}$ is adopted. Firstly, a sample with a specific value of $t_{up}$ is generated by MC simulations, and then obtain the efficiency of the detecting and event selection. Secondly, as for the every event in the data sample after event selection in the LHCb experiment, the duration time $t$ of $K^0(\bar{K}^0)$ in the laboratory frame can be obtained with the momentum and the flight length of $K^0(\bar{K}^0)$ in the laboratory frame. Thirdly, the duration time $t$ of $K^0(\bar{K}^0)$ in the rest frame of $K^0(\bar{K}^0)$ can be obtained by using the duration time $t$ of $K^0(\bar{K}^0)$ in the laboratory frame and the Lorentz transformations (between the laboratory frame and the rest frame of $B_c$ and between the rest frame of $B_c$ and the rest frame of $K^0(\bar{K}^0)$). Fourthly, to obtain the data sample which matchs with the theoretical prediction and MC simulations, the requirement that the duration time $t$ of $K^0(\bar{K}^0)$ in the rest frame of $K^0(\bar{K}^0)$ is less than the specific value of $t_{up}$ is applied in the data sample after event selection in the LHCb experiment. Obviously, this method is complicated, moreover, this method can't be performed perfectly for the $B_{c}^{+}\rightarrow B^{+} K^{0}+B^{+} {\bar{K}}^0\rightarrow B^{+} \pi^{+} e^{-} \bar{\nu}_{e}$ decays, this is because the final states contain neutrino, the energy and momentum of $B_{c}^{\pm}$ and $K^0(\bar{K}^0)$ mesons in the laboratory frame can't be determined accurately.
\item The events in which the flight length of $K^0(\bar{K}^0)$ is more than $2.5 m$ can also be detected by the tracking station (IT and OT) in the LHCb experiment, the efficiency of the detecting and event selection for these events is very low at LHCb. Meanwhile, when the duration time $t$ of $K^0(\bar{K}^0)$ is small, the $K^{0}(\bar{K}^0)\rightarrow \pi^{\pm} e^{\mp} {\nu}_{e}$ decays receive contributions from the $K_S^0$ decay and the $K_S^0-K_L^0$ interference, when the duration time $t$ of $K^0(\bar{K}^0)$ is large, the $K^{0}(\bar{K}^0)\rightarrow \pi^{\pm} e^{\mp} {\nu}_{e}$ decays receive contributions from the $K_L^0$ decay and the $K_S^0-K_L^0$ interference~\cite{DAmbrosio:2017klp,AlvesJunior:2018ldo}.
\item If we adopt the scenario that $t_{up}$ is infinity, the theoretical prediction is based on the calculations that include the contributions of all the duration time $t$ of $K^0(\bar{K}^0)$. In experiment, a sample with all the duration time $t$ of $K^0(\bar{K}^0)$ is generated by MC simulations, and then obtain the efficiency of the detecting and event selection. Meanwhile, the data sample after event selection also include the contributions of all the duration time $t$ of $K^0(\bar{K}^0)$ in the LHCb experiment.  Using the data sample after event selection in the LHCb experiment and the efficiency of the detecting and event selection, we can obtain the total event number which include the contribution of all the duration time $t$ of $K^0(\bar{K}^0)$ and match perfectly with the theoretical prediction.
\end{enumerate}
In this paper, we adopt the scenario that $t_{up}$ is infinity in order to match the theoretical prediction with the data sample in experiment.

Similarly, we can derive the time-independent decay widths for the $B_{c}^{-}\rightarrow B^{-} K^{0}+B^{-} {\bar{K}}^0\rightarrow B^{-} \pi^{-}  e^{+} \nu_{e}$ (hereinafter for brevity referred to as $B_{c}^{-}\rightarrow  B^{-} \pi^{-}  e^{+} \nu_{e}$) and $B_{c}^{\pm}\rightarrow B^{\pm} K^{0}+B^{\pm} {\bar{K}}^0\rightarrow B^{\pm} \pi^{\mp} e^{\pm} \nu_{e}$ (hereinafter for brevity referred to as $B_{c}^{\pm}\rightarrow  B^{\pm} \pi^{\mp} e^{\pm} \nu_{e}$) decays
\begin{align}
&\Gamma(B_{c}^{-}\rightarrow  B^{-} \pi^{-}  e^{+} \nu_{e})=\frac{G_F^2 f_K^2 \left(f_0(m_{K^0}^2)\right)^2}{288\pi m_{B_c^+}^3 }  f(m_{B_c^+},m_{B^+} ,m_{K^0})  (m_{B_c^+}^2-m_{B^+}^2)^2 \left|V_{cs}\right|^2 \left|V_{ud}\right|^2 \nonumber\\
&~\cdot\Gamma(K^{0}\rightarrow \pi^{-}  e^{+} \nu_{e})  \int_{0}^{\infty}\left|g_{+}(t)\right|^2 dt \left[ r_{wf}^{2} r_{g}\left|\frac{p}{q}\right|^2 +r_{wf} r_{sf} e^{i(\delta-\phi)}\frac{p}{q} + r_{wf} r_{sf} e^{-i(\delta-\phi)}\frac{p^{\ast}}{q^{\ast}}+1\right],\label{Eq:decaywidthbct2}\\
&\Gamma(B_{c}^{+}\rightarrow  B^{+} \pi^{-}  e^{+} \nu_{e})=\frac{G_F^2 f_K^2 \left(f_0(m_{K^0}^2)\right)^2}{288\pi m_{B_c^+}^3 }  f(m_{B_c^+},m_{B^+} ,m_{K^0})  (m_{B_c^+}^2-m_{B^+}^2)^2 \left|V_{cs}\right|^2 \left|V_{ud}\right|^2 \nonumber\\
&~\cdot\Gamma(K^{0}\rightarrow \pi^{-}  e^{+} \nu_{e})  \int_{0}^{\infty}\left|g_{+}(t)\right|^2 dt \left[ r_{wf}^{2} +r_{wf} r_{sf} e^{i(\delta-\phi)}\frac{p}{q} + r_{wf} r_{sf} e^{-i(\delta-\phi)}\frac{p^{\ast}}{q^{\ast}}+ r_{g}\left|\frac{p}{q}\right|^2\right],\label{Eq:decaywidthbct3}\\
&\Gamma(B_{c}^{-}\rightarrow  B^{-} \pi^{+} e^{-} \bar{\nu}_{e})=\frac{G_F^2 f_K^2 \left(f_0(m_{K^0}^2)\right)^2}{288\pi m_{B_c^+}^3 }  f(m_{B_c^+},m_{B^+} ,m_{K^0})  (m_{B_c^+}^2-m_{B^+}^2)^2 \left|V_{cs}\right|^2 \left|V_{ud}\right|^2 \nonumber\\
&~\cdot\Gamma(\bar{K}^{0}\rightarrow \pi^{+} e^{-} \bar{\nu}_{e})  \int_{0}^{\infty}\left|g_{+}(t)\right|^2 dt \left[ r_{wf}^{2} +r_{wf} r_{sf} e^{i(\delta+\phi)}\frac{q}{p} + r_{wf} r_{sf} e^{-i(\delta+\phi)}\frac{q^{\ast}}{p^{\ast}}+ r_{g}\left|\frac{q}{p}\right|^2\right],\label{Eq:decaywidthbct4}
\end{align}
where $\Gamma(K^{0}\rightarrow \pi^{-}  e^{+} \nu_{e})$ and $\Gamma(\bar{K}^{0}\rightarrow \pi^{+} e^{-} \bar{\nu}_{e}) $ are the decay widths for the $K^{0}\rightarrow \pi^{-}  e^{+} \nu_{e}$ and $\bar{K}^{0}\rightarrow \pi^{+} e^{-} \bar{\nu}_{e}$ decays, respectively. CPT invariance implies $\Gamma(K^{0}\rightarrow \pi^{-}  e^{+} \nu_{e})=\Gamma(\bar{K}^{0}\rightarrow \pi^{+} e^{-} \bar{\nu}_{e})$~\cite{Branco:1999fs}. Both $\Gamma(K^{0}\rightarrow \pi^{-}  e^{+} \nu_{e})$ and $\Gamma(\bar{K}^{0}\rightarrow \pi^{+} e^{-} \bar{\nu}_{e}) $ can be given by
\begin{align}
\Gamma(K^{0}\rightarrow \pi^{-}  e^{+} \nu_{e})=\Gamma(\bar{K}^{0}\rightarrow \pi^{+} e^{-} \bar{\nu}_{e})=\Gamma_S \cdot {\mathcal B}(K^0_S \rightarrow \pi^{\pm} e^{\mp} \nu_{e})=\Gamma_L \cdot {\mathcal B}(K^0_L \rightarrow \pi^{\pm} e^{\mp} \nu_{e}),\label{Eq:decaywidthkpienu}
\end{align}
where ${\mathcal B}(K^0_S \rightarrow \pi^{\pm} e^{\mp} \nu_{e})$ and ${\mathcal B}(K^0_L \rightarrow \pi^{\pm} e^{\mp} \nu_{e})$ are the branching ratios for the  $K^0_S \rightarrow \pi^{\pm} e^{\mp} \nu_{e}$ and $K^0_L \rightarrow \pi^{\pm} e^{\mp} \nu_{e}$ decays, respectively. Here, we note that ${\mathcal B}(K^0_{S} \rightarrow \pi^{\pm} e^{\mp} \nu_{e})$ is the summation of the branching ratios for the $K^0_S \rightarrow \pi^{+} e^{-} \bar{\nu}_{e}$ and $K^0_S \rightarrow \pi^{-} e^{+} \nu_{e}$ decays, ${\mathcal B}(K^0_L \rightarrow \pi^{\pm} e^{\mp} \nu_{e})$ is the summation of the branching ratios for the $K^0_L \rightarrow \pi^{+} e^{-} \bar{\nu}_{e}$ and $K^0_L \rightarrow \pi^{-} e^{+} \nu_{e}$ decays. Both ${\mathcal B}(K^0_{S} \rightarrow \pi^{\pm} e^{\mp} \nu_{e})$ and ${\mathcal B}(K^0_L \rightarrow \pi^{\pm} e^{\mp} \nu_{e})$ can be obtained from the Particle Data Group~\cite{ParticleDataGroup:2024cfk}. In order to characterize the different effects of CP violation, the Eq.(\ref{Eq:decaywidthbct2}) and Eq.(\ref{Eq:decaywidthbct4}) can also be written as
\begin{align}
&\Gamma(B_{c}^{-}\rightarrow  B^{-} \pi^{-}  e^{+} \nu_{e})=\frac{G_F^2 f_K^2 \left(f_0(m_{K^0}^2)\right)^2}{288\pi m_{B_c^+}^3 }  f(m_{B_c^+},m_{B^+} ,m_{K^0})  (m_{B_c^+}^2-m_{B^+}^2)^2 \left|V_{cs}\right|^2 \left|V_{ud}\right|^2 \nonumber\\
&~\cdot\Gamma(K^{0}\rightarrow \pi^{-}  e^{+} \nu_{e})  \int_{0}^{\infty}\left|g_{+}(t)\right|^2 dt \left[ r_{wf}^{2} r_{g}\left|\frac{p}{q}\right|^2 +r_{wf} r_{sf} e^{i(\delta-\phi)}\frac{q}{p} + r_{wf} r_{sf} e^{-i(\delta-\phi)}\frac{q^{\ast}}{p^{\ast}}+1\right.\nonumber\\
&~~~~~~~~~~~~~~~~~~~~~~~~~~~~~~~~~~~~~~~~~~\left.-r_{wf} r_{sf} e^{i(\delta-\phi)}\left(\frac{q}{p}-\frac{p}{q}\right)-r_{wf} r_{sf} e^{-i(\delta-\phi)}\left(\frac{q^{\ast}}{p^{\ast}}-\frac{p^{\ast}}{q^{\ast}}\right)\right],\label{Eq:redecaywidthbct1}\\
&\Gamma(B_{c}^{-}\rightarrow  B^{-} \pi^{+} e^{-} \bar{\nu}_{e})=\frac{G_F^2 f_K^2 \left(f_0(m_{K^0}^2)\right)^2}{288\pi m_{B_c^+}^3 }  f(m_{B_c^+},m_{B^+} ,m_{K^0})  (m_{B_c^+}^2-m_{B^+}^2)^2 \left|V_{cs}\right|^2 \left|V_{ud}\right|^2 \nonumber\\
&~\cdot\Gamma(\bar{K}^{0}\rightarrow \pi^{+} e^{-} \bar{\nu}_{e})  \int_{0}^{\infty}\left|g_{+}(t)\right|^2 dt \left[ r_{wf}^{2} +r_{wf} r_{sf} e^{i(\delta+\phi)}\frac{p}{q} + r_{wf} r_{sf} e^{-i(\delta+\phi)}\frac{p^{\ast}}{q^{\ast}}+ r_{g}\left|\frac{q}{p}\right|^2\right.\nonumber\\
&~~~~~~~~~~~~~~~~~~~~~~~~~~~~~~~~~~~~~~~~~~\left.-r_{wf} r_{sf} e^{i(\delta+\phi)}\left(\frac{p}{q}-\frac{q}{p}\right)-r_{wf} r_{sf} e^{-i(\delta+\phi)}\left(\frac{p^{\ast}}{q^{\ast}}-\frac{q^{\ast}}{p^{\ast}}\right)\right],\label{Eq:redecaywidthbct2}
\end{align}
where the second and third terms in the square bracket of the above equations corresponding to the direct CP violation effects, the last two terms in the square bracket of the above equations corresponding the new CP violation effects, which will discussed in the next section.

Making use of Eqs.(\ref{Eq:decaywidthbct1}), (\ref{Eq:decaywidthbct2})-(\ref{Eq:decaywidthbct4}) and multiplying the mean life of $B_{c}$ meson, we can obtain the branching ratios for the $B_{c}^{+}\rightarrow  B^{+} \pi^{\pm} e^{\mp} \nu_{e}$ and $B_{c}^{-}\rightarrow  B^{-} \pi^{\pm} e^{\mp} \nu_{e}$ decays.
\section{the asymmetries in $B_{c}\rightarrow B \pi^{\pm} e^{\mp} \nu_{e}$ decays }
\label{sec:cpviolationcal}
Basing on the partial decay widths for the $B_{c}^{+}\rightarrow  B^{+} \pi^{\pm} e^{\mp} \nu_{e}$ and $B_{c}^{-}\rightarrow  B^{-} \pi^{\pm} e^{\mp} \nu_{e}$ decays derived in section~\ref{sec:decaywidth}, we can proceed to calculate the asymmetries in these decays.
\subsection{CP asymmetry between the $B_{c}^{+} \rightarrow B^{+} \pi^{+} e^{-} \bar{\nu}_{e}$ and $B_{c}^{-} \rightarrow B^{-} \pi^{-}  e^{+} \nu_{e}$ decays }
\label{sec:cpcalacppm}
In cascade decay chains $B_{c}^{\pm}\rightarrow  B^{\pm} \pi^{\pm} e^{\mp} \nu_{e}\rightarrow f_{B^{\pm}} \pi^{\pm} e^{\mp} \nu_{e}$, the time-independent CP violation observable is defined as
\begin{align}
{\mathcal A}_{CP}^{pm}=\frac{\Gamma(B_c^+\rightarrow B^{+} \pi^{+} e^{-} \bar{\nu}_{e})\cdot\Gamma(B^{+}\rightarrow f_{B})-\Gamma(B_c^-\rightarrow B^{-} \pi^{-} e^{+} \nu_{e})\cdot\Gamma(B^{-}\rightarrow \bar{f}_{B})}{\Gamma(B_c^+\rightarrow B^{+} \pi^{+} e^{-} \bar{\nu}_{e})\cdot\Gamma(B^{+}\rightarrow f_{B})+\Gamma(B_c^-\rightarrow B^{-} \pi^{-}  e^{+} \nu_{e})\cdot\Gamma(B^{-}\rightarrow \bar{f}_{B})},\label{Eq:cpasymmacppm1}
\end{align}
where $f_{B}$ denotes the final state from the decay of the $B^+$ meson, $\bar{f}_{B}$ is the CP-conjugate state of $f_{B}$, $\Gamma(B^{+}\rightarrow f_{B})$ and $\Gamma(B^{-}\rightarrow \bar{f}_{B})$ are the partial decay widths of the $B^{+}\rightarrow f_{B}$ and $B^{-}\rightarrow \bar{f}_{B}$ decays respectively. With the Eqs.(\ref{Eq:bianhuanrelation1}) and (\ref{Eq:bianhuanrelation2}) listed in Appendix~\ref{sec:appendixcpcal}, we can obtain
\begin{align}
&{\mathcal A}_{CP}^{pm}={\mathcal A}_{CP}\left(B_c^{\pm}\rightarrow B^{\pm} \pi^{\pm} e^{\mp} \nu_{e}\right)\cdot \left[1-{\mathcal A}_{CP}\left(B\rightarrow f_{B}\right)+o\left({\mathcal A}_{CP}\left(B\rightarrow f_{B}\right)\right)\right]\nonumber\\
&~~~~~~~~+{\mathcal A}_{CP}\left(B\rightarrow f_{B}\right)\cdot \left[1+{\mathcal A}_{CP}\left(B_c^{\pm}\rightarrow B^{\pm} \pi^{\pm} e^{\mp} \nu_{e}\right)+o\left({\mathcal A}_{CP}\left(B_c^{\pm}\rightarrow B^{\pm} \pi^{\pm} e^{\mp} \nu_{e}\right)\right)\right]\nonumber\\
&~~~~~~={\mathcal A}_{CP}\left(B_c^{\pm}\rightarrow B^{\pm} \pi^{\pm} e^{\mp} \nu_{e}\right)\cdot \left[1+o\left({\mathcal A}_{CP}\left(B\rightarrow f_{B}\right)\right)\right]\nonumber\\
&~~~~~~~~~~~~~~~~~~~~~~~~~~~~~~~~~~~~~~~~~~~+{\mathcal A}_{CP}\left(B\rightarrow f_{B}\right)\cdot \left[1+o\left({\mathcal A}_{CP}\left(B_c^{\pm}\rightarrow B^{\pm} \pi^{\pm} e^{\mp} \nu_{e}\right)\right)\right],\label{Eq:cpasymmacppm2}
\end{align}
where ${\mathcal A}_{CP}\left(B_c^{\pm}\rightarrow B^{\pm} \pi^{\pm} e^{\mp} \nu_{e}\right)$ and ${\mathcal A}_{CP}\left(B\rightarrow f_{B}\right)$ denote the CP asymmetry in $B_c^+\rightarrow B^{+} \pi^{+} e^{-} \bar{\nu}_{e}$ and $B_c^-\rightarrow B^{-} \pi^{-}  e^{+} \nu_{e}$ decays and the CP asymmetry in $B^+\rightarrow f_{B}$ and $B^-\rightarrow \bar{f}_{B}$ decays, respectively,
\begin{align}
&{\mathcal A}_{CP}\left(B\rightarrow f_{B}\right)=\frac{\Gamma(B^{+}\rightarrow f_{B})-\Gamma(B^{-}\rightarrow \bar{f}_{B})}{\Gamma(B^{+}\rightarrow f_{B})+\Gamma(B^{-}\rightarrow \bar{f}_{B})},\label{Eq:defacpbpmfb}\\
&{\mathcal A}_{CP}\left(B_c^{\pm}\rightarrow B^{\pm} \pi^{\pm} e^{\mp} \nu_{e}\right)=\frac{\Gamma(B_c^+\rightarrow B^{+} \pi^{+} e^{-} \bar{\nu}_{e})-\Gamma(B_c^-\rightarrow B^{-} \pi^{-}  e^{+} \nu_{e})}{\Gamma(B_c^+\rightarrow B^{+} \pi^{+} e^{-} \bar{\nu}_{e})+\Gamma(B_c^-\rightarrow B^{-} \pi^{-}  e^{+} \nu_{e})},\label{Eq:defacpbcpmbpmpi}
\end{align}
$o\left({\mathcal A}_{CP}\left(B_c^{\pm}\rightarrow B^{\pm} \pi^{\pm} e^{\mp} \nu_{e}\right)\right)$ and $o\left({\mathcal A}_{CP}\left(B_c^{\pm}\rightarrow B^{\pm} \pi^{\pm} e^{\mp} \nu_{e}\right)\right)$ in Eq.(\ref{Eq:cpasymmacppm2}) denote the high-order small quantities. 

The CP violation effects in the nonleptonic decays of the $B^{\pm}$ can reach the value larger than $0.1$~\cite{ParticleDataGroup:2024cfk}, so they can't be neglected. Fortunately, we only investigate the semileptonic decays of the $B^{\pm}$ mesons in this paper,  the direct CP violations in these decays don't occur within the Standard Model~\cite{D0:2016xvr}. However, ${\mathcal A}_{CP}\left(B\rightarrow f_{B}\right)$ should also receive the contributions from the CP violations in charm meson decays for the semileptonic decays of the $B^{\pm}$ to charmed meson, for example, the CP violation in $B^{\pm}\rightarrow \bar{D}^0 l^{\pm}\nu_l\rightarrow f_{\bar{D}} l^{\pm}\nu_l$ decay chains is given by
\begin{align}
&{\mathcal A}_{CP}\left(B\rightarrow f_{B}\right)=\frac{\Gamma(B^{+}\rightarrow  \bar{D}^0 l^+ \nu_l)-\Gamma(B^{-}\rightarrow  D^0 l^- \bar{\nu}_l)}{\Gamma(B^{+}\rightarrow  \bar{D}^0 l^+ \nu_l)+\Gamma(B^{-}\rightarrow  D^0 l^- \bar{\nu}_l)}+\frac{\Gamma(\bar{D}^0\rightarrow f_{\bar{D}})-\Gamma(D^0\rightarrow \bar{f}_{D})}{\Gamma(\bar{D}^0\rightarrow f_{\bar{D}})+\Gamma(D^0\rightarrow \bar{f}_{D})},\label{Eq:cpbplusdzlvl}
\end{align}
where $\left(\Gamma(\bar{D}^0\rightarrow f_{\bar{D}})-\Gamma(D^0\rightarrow \bar{f}_{D})\right)/\left(\Gamma(\bar{D}^0\rightarrow f_{\bar{D}})+\Gamma(D^0\rightarrow \bar{f}_{D})\right)$ is the CP asymmetry in $\bar{D}^0\rightarrow f_{\bar{D}}$ and $D^0\rightarrow \bar{f}_{D}$ decays, $f_{\bar{D}}$ denotes the final state from the decay of the $\bar{D}^0$ meson, $\bar{f}_{D}$ is the CP-conjugate state of $f_{\bar{D}}$. In this paper, the
$\bar{D}^0\rightarrow K^+ e^- \bar{\nu}_e$, $\bar{D}^0\rightarrow K^+ \mu^- \bar{\nu}_\mu$, $\bar{D}^0\rightarrow K^+ \pi^-$, $\bar{D}^0\rightarrow K^+ \pi^- \pi^0$, $\bar{D}^0\rightarrow K^+ \pi^- \pi^- \pi^+$, $\bar{D}^0\rightarrow K^+ \pi^- \pi^- \pi^+ \pi^0$ and $\bar{D}^0\rightarrow K^+ \pi^- \pi^0 \pi^0$ decays are involved. For the semileptonic decays of the $D^0$ meson, there is no CP violation in the Standard Model. For the nonleptonic $D^0$ decays listed above, the CP violation hasn't been discovered in experiment. With the measurement of the CP asymmetry in $D^0\rightarrow \pi^+ \pi^-$ decay, the branching ratio for the $D^0\rightarrow \pi^+ \pi^-$ decay and the large branching ratio for
the  nonleptonic $D^0$ decays listed above, the orders of the CP violations in the nonleptonic $D^0$ decays listed above can be estimated to be $\mathcal{O}(10^{-5})$ or smaller, which can be neglected safely~\cite{ParticleDataGroup:2024cfk,LHCb:2022lry,LHCb:2025kch,Kmiec:2025auz}. In summary, the CP asymmetry in the semileptonic decays of the $B^{\pm}$ is
\begin{align}
&{\mathcal A}_{CP}\left(B\rightarrow f_{B}\right)=\frac{\Gamma(B^{+}\rightarrow f_{B})-\Gamma(B^{-}\rightarrow \bar{f}_{B})}{\Gamma(B^{+}\rightarrow f_{B})+\Gamma(B^{-}\rightarrow \bar{f}_{B})}\approx 0,\label{Eq:cpbpbmfbnum}
\end{align}
combining Eq.(\ref{Eq:cpasymmacppm2}), Eq.(\ref{Eq:defacpbpmfb}), Eq.(\ref{Eq:defacpbcpmbpmpi}) and Eq.(\ref{Eq:cpbpbmfbnum}), we can obtain
\begin{align}
{\mathcal A}_{CP}^{pm}=\frac{\Gamma(B_c^+\rightarrow B^{+} \pi^{+} e^{-} \bar{\nu}_{e})-\Gamma(B_c^-\rightarrow B^{-} \pi^{-}  e^{+} \nu_{e})}{\Gamma(B_c^+\rightarrow B^{+} \pi^{+} e^{-} \bar{\nu}_{e})+\Gamma(B_c^-\rightarrow B^{-} \pi^{-}  e^{+} \nu_{e})}.\label{Eq:cpasymmacppm3}
\end{align}
By combining Eq.(\ref{Eq:ksdefinition3}), Eq.(\ref{Eq:decaywidthbct1}), Eq.(\ref{Eq:definrwfrsfrg}), Eq.(\ref{Eq:gpmtintgrate3}), Eq.(\ref{Eq:redecaywidthbct1}) and Eq.(\ref{Eq:cpasymmacppm3}), we can derive
\begin{align}
&{\mathcal A}_{CP}^{pm}={\mathcal A}_{CP,mix}^{pm}+{\mathcal A}_{CP,dir}^{pm}+{\mathcal A}_{CP,int}^{pm},\label{Eq:cpasymmacppm31}\\
&{\mathcal A}_{CP,mix}^{pm}=\frac{r_{wf}^{2}\cdot r_{g}\cdot \left(\left|\frac{q}{p}\right|^2-\left|\frac{p}{q}\right|^2\right)}{r_{wf}^{2}\cdot r_{g}\cdot \left(\left|\frac{q}{p}\right|^2 + \left|\frac{p}{q}\right|^2\right)+2r_{wf} \cdot r_{sf}\cdot Re\left(e^{-i\phi}\left(\frac{q^{\ast}}{p^{\ast}}e^{-i\delta}+\frac{p}{q}e^{i\delta}\right)\right)+2},\label{Eq:cpasymmacppm32}\\
&{\mathcal A}_{CP,dir}^{pm}=\frac{-4 r_{wf}\cdot r_{sf}\cdot \left|\frac{q}{p}\right|\cdot \sin\phi\cdot\sin\delta_{S}^{pm} }{r_{wf}^{2}\cdot r_{g}\cdot \left(\left|\frac{q}{p}\right|^2 + \left|\frac{p}{q}\right|^2\right)+2r_{wf} \cdot r_{sf}\cdot Re\left(e^{-i\phi}\left(\frac{q^{\ast}}{p^{\ast}}e^{-i\delta}+\frac{p}{q}e^{i\delta}\right)\right)+2},\label{Eq:cpasymmacppm33}\\
&{\mathcal A}_{CP,int}^{pm}=\frac{2 r_{wf}\cdot r_{sf}\cdot Re\left(e^{-i\left(\phi-\delta\right)} \left(\frac{q}{p}-\frac{p}{q}\right)\right)}{r_{wf}^{2}\cdot r_{g}\cdot \left(\left|\frac{q}{p}\right|^2 + \left|\frac{p}{q}\right|^2\right)+2r_{wf} \cdot r_{sf}\cdot Re\left(e^{-i\phi}\left(\frac{q^{\ast}}{p^{\ast}}e^{-i\delta}+\frac{p}{q}e^{i\delta}\right)\right)+2},\label{Eq:cpasymmacppm34}
\end{align}
where ${\mathcal A}_{CP,mix}^{pm}$ denotes the indirect CP violation in kaon mixing~\cite{Grossman:2011zk,Yu:2017oky}, which is induced by the difference between the time-evolutions of an initially $(t=0)$ pure $K^0$ and $\bar{K}^0$ as shown in Eq.(\ref{Eq:kzphysdef}) and Eq.(\ref{Eq:kbphysdef}). ${\mathcal A}_{CP,dir}^{pm}$ denotes the direct CP violation originating from the interference between the Cabibbo-favored (CF) and doubly Cabibbo-suppressed (DCS) amplitudes. The parameter $\delta_{S}^{pm}$ in Eq.(\ref{Eq:cpasymmacppm33}) denotes the strong phase difference of the direct CP asymmetry ${\mathcal A}_{CP,dir}^{pm}$,  which is given by
\begin{align}
&\delta_{S}^{pm}=arg\left(\frac{q}{p}\right)+\delta=-arctan\left(\frac{2Im(\epsilon)}{1-|\epsilon|^2}\right)-arctan\left(\frac{2\Delta m}{\Gamma_S -\Gamma_L}\cdot \frac{\Gamma_S \Gamma_L}{\Gamma^2+(\Delta m)^2}\right).\label{Eq:cpasymmacppmstrph}
\end{align}
From the above equation, we can see that the strong phase difference of the direct CP asymmetry ${\mathcal A}_{CP,dir}^{pm}$ is associated with the phases of $q/p$ and $\int_{0}^{\infty}g_{-}(t) \left(g_{+}(t)\right)^{\ast} dt$, i.e., the strong phase difference of the direct CP asymmetry arising from $K^0-\bar{K}^0$ mixing parameters, and thus preventing pollution from strong dynamics. ${\mathcal A}_{CP,int}^{pm}$ represents the new CP violation effect (hereinafter for brevity referenced as the CP violation in the interference), which originates from the interference between the amplitude of the $B_{c}^{-}\rightarrow B^{-} K^{0}\rightarrow B^{-} \pi^{-}  e^{+} \nu_{e}$ decay with the difference between the mixing effect of $K^{0}\rightarrow \bar{K}^0$ and that of $\bar{K}^0\rightarrow K^{0}$, which can be derived from Eqs.(\ref{Eq:kzphysdef}), (\ref{Eq:kbphysdef}), (\ref{Eq:decaywidthbct1}) and (\ref{Eq:redecaywidthbct1}). Combining the Eq.(\ref{Eq:ksdefinition3}) and Eq.(\ref{Eq:cpasymmacppm34}), we can obtain that
\begin{align}
&{\mathcal A}_{CP,int}^{pm}=\frac{-8 r_{wf}\cdot r_{sf}\cdot \left(Re\epsilon \cdot\cos\left(\phi-\delta\right) +Im\epsilon\cdot\sin\left(\phi-\delta\right) \right)}{r_{wf}^{2}\cdot r_{g}\cdot \left(\left|\frac{q}{p}\right|^2 + \left|\frac{p}{q}\right|^2\right)+2r_{wf} \cdot r_{sf}\cdot Re\left(e^{-i\phi}\left(\frac{q^{\ast}}{p^{\ast}}e^{-i\delta}+\frac{p}{q}e^{i\delta}\right)\right)+2},\label{Eq:cpasymmacppm35}
\end{align}
obviously, ${\mathcal A}_{CP,int}^{pm}$ involves the parameter $r_{wf}$ and the weak phase difference $\phi$, which relate to the CKM matrix elements.
\subsection{CP asymmetry between the $B_{c}^{+} \rightarrow B^{+} \pi^{-} e^{+} \nu_{e}$ and $B_{c}^{-} \rightarrow B^{-} \pi^{+}  e^{-} \bar{\nu}_{e}$ decays }
\label{sec:cpcalacpmp}
In cascade decay chains $B_{c}^{\pm}\rightarrow  B^{\pm} \pi^{\mp} e^{\pm} \nu_{e}\rightarrow f_{B^{\pm}} \pi^{\mp} e^{\pm} \nu_{e}$, the time-independent CP violation observable is defined as
\begin{align}
{\mathcal A}_{CP}^{mp}=\frac{\Gamma(B_c^+\rightarrow B^{+} \pi^{-} e^{+} \nu_{e})\cdot\Gamma(B^{+}\rightarrow f_{B})-\Gamma(B_c^-\rightarrow B^{-} \pi^{+} e^{-} \bar{\nu}_{e})\cdot\Gamma(B^{-}\rightarrow \bar{f}_{B})}{\Gamma(B_c^+\rightarrow B^{+} \pi^{-} e^{+} \nu_{e})\cdot\Gamma(B^{+}\rightarrow f_{B})+\Gamma(B_c^-\rightarrow B^{-} \pi^{+} e^{-} \bar{\nu}_{e})\cdot\Gamma(B^{-}\rightarrow \bar{f}_{B})},\label{Eq:cpasymmacpmp1}
\end{align}
using the same method as that for deriving the expression of ${\mathcal A}_{CP}^{pm}$ and the Eqs.(\ref{Eq:bianhuanrelation1})-(\ref{Eq:bianhuanrelation2}), we can derive
\begin{align}
{\mathcal A}_{CP}^{mp}=\frac{\Gamma(B_c^+\rightarrow B^{+} \pi^{-} e^{+} \nu_{e})-\Gamma(B_c^-\rightarrow B^{-} \pi^{+} e^{-} \bar{\nu}_{e})}{\Gamma(B_c^+\rightarrow B^{+} \pi^{-} e^{+} \nu_{e})+\Gamma(B_c^-\rightarrow B^{-} \pi^{+} e^{-} \bar{\nu}_{e})}.\label{Eq:cpasymmacpmp2}
\end{align}
Using the Eq.(\ref{Eq:ksdefinition3}), Eq.(\ref{Eq:decaywidthbct3}), Eq.(\ref{Eq:redecaywidthbct2}) and Eq.(\ref{Eq:cpasymmacpmp2}), we can obtain
\begin{align}
&{\mathcal A}_{CP}^{mp}={\mathcal A}_{CP,mix}^{mp}+{\mathcal A}_{CP,dir}^{mp}+{\mathcal A}_{CP,int}^{mp},\label{Eq:cpasymmacpmp31}\\
&{\mathcal A}_{CP,mix}^{mp}=\frac{ r_{g}\cdot \left(\left|\frac{p}{q}\right|^2-\left|\frac{q}{p}\right|^2\right)}{2 r_{wf}^{2} +2r_{wf} \cdot r_{sf}\cdot Re\left(e^{-i\phi}\left(\frac{p}{q}e^{i\delta}+\frac{q^{\ast}}{p^{\ast}}e^{-i\delta}\right)\right) + r_{g}\cdot \left(\left|\frac{p}{q}\right|^2 + \left|\frac{q}{p}\right|^2\right)},\label{Eq:cpasymmacpmp32}\\
&{\mathcal A}_{CP,dir}^{mp}=\frac{4 r_{wf}\cdot r_{sf}\cdot \left|\frac{p}{q}\right|\cdot \sin\phi\cdot \sin\delta_{S}^{mp} }{2 r_{wf}^{2} +2r_{wf} \cdot r_{sf}\cdot Re\left(e^{-i\phi}\left(\frac{p}{q}e^{i\delta}+\frac{q^{\ast}}{p^{\ast}}e^{-i\delta}\right)\right) + r_{g}\cdot \left(\left|\frac{p}{q}\right|^2 + \left|\frac{q}{p}\right|^2\right)},\label{Eq:cpasymmacpmp33}\\
&{\mathcal A}_{CP,int}^{mp}=\frac{2 r_{wf}\cdot r_{sf}\cdot Re\left(e^{i\left(\phi+\delta\right)} \left(\frac{p}{q}-\frac{q}{p}\right)\right)}{2 r_{wf}^{2} +2r_{wf} \cdot r_{sf}\cdot Re\left(e^{-i\phi}\left(\frac{p}{q}e^{i\delta}+\frac{q^{\ast}}{p^{\ast}}e^{-i\delta}\right)\right) + r_{g}\cdot \left(\left|\frac{p}{q}\right|^2 + \left|\frac{q}{p}\right|^2\right)},\label{Eq:cpasymmacpmp34}
\end{align}
where ${\mathcal A}_{CP,mix}^{mp}$, ${\mathcal A}_{CP,dir}^{mp}$ and ${\mathcal A}_{CP,int}^{mp}$ denote the indirect CP violation in kaon mixing, the direct CP violation and the CP violation in the interference, respectively. The parameter $\delta_{S}^{mp}$ in Eq.(\ref{Eq:cpasymmacpmp33}) denotes the strong phase difference in the direct CP violation ${\mathcal A}_{CP,dir}^{mp}$, which is given by
\begin{align}
&\delta_{S}^{mp}=arg\left(\frac{p}{q}\right)+\delta=arctan\left(\frac{2Im(\epsilon)}{1-|\epsilon|^2}\right)-arctan\left(\frac{2\Delta m}{\Gamma_S -\Gamma_L}\cdot \frac{\Gamma_S \Gamma_L}{\Gamma^2+(\Delta m)^2}\right).\label{Eq:cpasymmacpmpstrph}
\end{align}
As can be seen in the above equation, the strong phase difference of the direct CP asymmetry arising from kaon mixing parameters, and thus free of strong pollution. Using the Eq.(\ref{Eq:ksdefinition3}) and Eq.(\ref{Eq:cpasymmacpmp34}), we can rewrite the CP violation in the interference ${\mathcal A}_{CP,int}^{mp}$ as following
\begin{align}
&{\mathcal A}_{CP,int}^{mp}=\frac{8 r_{wf}\cdot r_{sf}\cdot \left(Re\epsilon \cdot\cos\left(\phi+\delta\right) -Im\epsilon\cdot\sin\left(\phi+\delta\right) \right)}{2 r_{wf}^{2} +2r_{wf} \cdot r_{sf}\cdot Re\left(e^{-i\phi}\left(\frac{p}{q}e^{i\delta}+\frac{q^{\ast}}{p^{\ast}}e^{-i\delta}\right)\right) + r_{g}\cdot \left(\left|\frac{p}{q}\right|^2 + \left|\frac{q}{p}\right|^2\right)},\label{Eq:cpasymmacpmp35}
\end{align}
similarly, ${\mathcal A}_{CP,int}^{mp}$ involves the parameter $r_{wf}$ and the weak phase differenc $\phi$, which are determined by the CKM matrix elements.
\subsection{the asymmetries between the $B_{c}^+ \rightarrow B^+ \pi^{\pm} e^{\mp} \nu_{e}$ decays and their  partially conjugated decays $B_{c}^- \rightarrow B^- \pi^{\pm} e^{\mp} \nu_{e}$ }
\label{sec:asycalappmm}
Beside the CP asymmetries in $B_{c}^{+} \rightarrow B^{+} \pi^{\pm} e^{\mp} \nu_{e}$ and their conjugated decays, we can also define another two asymmetry observables, which are given by
\begin{align}
&{\mathcal A}_{CP}^{pp}=\frac{\Gamma(B_c^+\rightarrow B^{+} \pi^{+} e^{-} \bar{\nu}_{e})\cdot\Gamma(B^{+}\rightarrow f_{B})-\Gamma(B_c^-\rightarrow B^{-} \pi^{+} e^{-} \bar{\nu}_{e})\cdot\Gamma(B^{-}\rightarrow \bar{f}_{B})}{\Gamma(B_c^+\rightarrow B^{+} \pi^{+} e^{-} \bar{\nu}_{e})\cdot\Gamma(B^{+}\rightarrow f_{B})+\Gamma(B_c^-\rightarrow B^{-} \pi^{+} e^{-} \bar{\nu}_{e})\cdot\Gamma(B^{-}\rightarrow \bar{f}_{B})},\label{Eq:cpasymmacppp1}
\end{align}
and
\begin{align}
&{\mathcal A}_{CP}^{mm}=\frac{\Gamma(B_c^+\rightarrow B^{+} \pi^{-} e^{+} \nu_{e})\cdot\Gamma(B^{+}\rightarrow f_{B})-\Gamma(B_c^-\rightarrow B^{-} \pi^{-} e^{+} \nu_{e})\cdot\Gamma(B^{-}\rightarrow \bar{f}_{B})}{\Gamma(B_c^+\rightarrow B^{+} \pi^{-} e^{+} \nu_{e})\cdot\Gamma(B^{+}\rightarrow f_{B})+\Gamma(B_c^-\rightarrow B^{-} \pi^{-} e^{+} \nu_{e})\cdot\Gamma(B^{-}\rightarrow \bar{f}_{B})}.\label{Eq:cpasymmacpmm1}
\end{align}
Using the Eqs.(\ref{Eq:bianhuanrelappmm1})-(\ref{Eq:bianhuanrelappmm2}) and employing the same methods in subsection~\ref{sec:cpcalacppm}, we can simplify Eq.(\ref{Eq:cpasymmacppp1}) and Eq.(\ref{Eq:cpasymmacpmm1}) as
\begin{align}
&{\mathcal A}_{CP}^{pp}=\frac{\Gamma(B_c^+\rightarrow B^{+} \pi^{+} e^{-} \bar{\nu}_{e})-\Gamma(B_c^-\rightarrow B^{-} \pi^{+} e^{-} \bar{\nu}_{e})}{\Gamma(B_c^+\rightarrow B^{+} \pi^{+} e^{-} \bar{\nu}_{e})+\Gamma(B_c^-\rightarrow B^{-} \pi^{+} e^{-} \bar{\nu}_{e})},\label{Eq:cpasymmacppp2}\\
&{\mathcal A}_{CP}^{mm}=\frac{\Gamma(B_c^+\rightarrow B^{+} \pi^{-} e^{+} \nu_{e})-\Gamma(B_c^-\rightarrow B^{-} \pi^{-} e^{+} \nu_{e})}{\Gamma(B_c^+\rightarrow B^{+} \pi^{-} e^{+} \nu_{e})+\Gamma(B_c^-\rightarrow B^{-} \pi^{-} e^{+} \nu_{e})}.\label{Eq:cpasymmacpmm2}
\end{align}
By substituting Eqs.(\ref{Eq:decaywidthbct1}), (\ref{Eq:decaywidthbct2}), (\ref{Eq:decaywidthbct3}) and (\ref{Eq:decaywidthbct4}) into Eqs.(\ref{Eq:cpasymmacppp2})-(\ref{Eq:cpasymmacpmm2}), we can obtain
\begin{align}
&{\mathcal A}_{CP}^{pp}=\frac{\left|p\right|^2-\left|q\right|^2 \cdot r_g}{\left|p\right|^2+\left|q\right|^2 \cdot r_g} =\frac{\left|p\right|^2-\left|q\right|^2 +\left|q\right|^2\cdot \left(1- r_g\right)}{\left|p\right|^2+\left|q\right|^2 \cdot r_g},\label{Eq:cpasymmacppp3}\\
&{\mathcal A}_{CP}^{mm}=-\frac{\left|q\right|^2-\left|p\right|^2 \cdot r_g}{\left|q\right|^2+\left|p\right|^2 \cdot r_g}=\frac{\left|p\right|^2-\left|q\right|^2-\left|p\right|^2\cdot \left(1- r_g\right)}{\left|q\right|^2+\left|p\right|^2 \cdot r_g}.\label{Eq:cpasymmacpmm3}
\end{align}
From the above equations, we can see that the dominant contributions to the asymmetry observables ${\mathcal A}_{CP}^{pp}$ and ${\mathcal A}_{CP}^{mm}$ arising from $K^0-{\bar{K}}^0$ mixing, the reasons are as follows: in the asymmetry observable ${\mathcal A}_{CP}^{pp}$, the involved decays is not conjugated with each other, but have the same final states $(\pi^{+} e^{-} \bar{\nu}_{e})$ of the neutral kaon, so not only the weak phase difference but also the strong phase difference between the Cabibbo-favored and doubly Cabibbo-suppressed amplitudes of the decay process $B_c^+\rightarrow B^{+} \pi^{+} e^{-} \bar{\nu}_{e}$ are opposite to that of the decay process $B_c^-\rightarrow B^{-} \pi^{+} e^{-} \bar{\nu}_{e}$, which can be seen in Eq.(\ref{Eq:ampjtbcpmpipmemp1}) and Eq.(\ref{Eq:ampjtbcpmpipmemp4}), as a result, the interference between the Cabibbo-favored and doubly Cabibbo-suppressed amplitudes of the decay process $B_c^+\rightarrow B^{+} \pi^{+} e^{-} \bar{\nu}_{e}$ is the same as that of the $B_c^-\rightarrow B^{-} \pi^{+} e^{-} \bar{\nu}_{e}$ decay, as shown in Eq.(\ref{Eq:decaywidthbct1}) and Eq.(\ref{Eq:decaywidthbct4}), so there is no direct CP violation. Meanwhile, because the decays in ${\mathcal A}_{CP}^{pp}$ have the same final states $(\pi^{+} e^{-} \bar{\nu}_{e})$ of the neutral kaon, only the mixing effect of $K^{0}\rightarrow \bar{K}^0$ is involved, so ${\mathcal A}_{CP}^{pp}$ have no CP violation in the interference, which originates from the interference between the amplitude of the $B_{c}^{-}\rightarrow B^{-} K^{0} (\bar{K}^0)\rightarrow B^{-} \pi^{-}  e^{+} \nu_{e} (\pi^{+}  e^{-} \bar{\nu}_{e})$ decay with the difference between the mixing effect of $K^{0}\rightarrow \bar{K}^0$ and that of $\bar{K}^0\rightarrow K^{0}$. The situation in the asymmetry observable ${\mathcal A}_{CP}^{mm}$ is the same as that in the asymmetry observable ${\mathcal A}_{CP}^{pp}$.
\section{Numerical calculation}
\label{sec:numberres}
\subsection{Input parameters}
\label{sec:inputparameters}
Using the theoretical expressions for the branching ratios and the asymmetry observables derived in section~\ref{sec:decaywidth} and section~\ref{sec:cpviolationcal}, we are able to calculate the observables numerically. Firstly, we collect the input parameters used in this work as below~\cite{ExtendedTwistedMass:2020tvp,ParticleDataGroup:2024cfk}
\begin{align}
&m_{B_c^+}=6.274\text{GeV}, &&m_{B^+}=5.279 \text{GeV},\nonumber\\
&\Delta m=\left(3.484\pm0.006\right)\times 10^{-15}\text{GeV},  &&G_{F}=1.166\times 10^{-5}\text{GeV}^{-2},\nonumber\\
&\Gamma_{S}=(7.351\pm0.003)\times 10^{-15}\text{GeV},  &&\Gamma_L=(1.287\pm0.005)\times 10^{-17} \text{GeV},\nonumber\\
&m_{D^*}=2.007\text{GeV},  && m_{D_0^{*}}=(2.343\pm0.010)\text{GeV}, \label{Eq:valparameters}\\
&\tau_{B_c}=\left(0.510\pm0.009\right)\times 10^{-12}\text{s},  &&f_{K}=(0.154\pm0.002)\text{GeV},\nonumber\\
&Re(\epsilon)=(1.66\pm0.02)\times 10^{-3},  &&Im(\epsilon)=(1.57\pm0.02)\times 10^{-3},\nonumber\\
&m_{K^0}=0.498\text{GeV},  && {\mathcal B}(K^0_S \rightarrow \pi^{\pm} e^{\mp} \nu_{e})=(7.14\pm0.06)\times 10^{-4}.\nonumber
\end{align}
In order to see physics more transparently, we use the Wolfenstein parametrization of the CKM matrix elements~\cite{ParticleDataGroup:2024cfk,Wolfenstein:1983yz,Ahn:2011fg,Buras:1998raa}
\begin{align}
&V_{ud}=\sqrt{1-\lambda^2}\sqrt{1-A^2\lambda^6(\rho^2+\eta^2)}, ~~~ &V_{cd}&=-\lambda\sqrt{1-A^2\lambda^4}-A^2\lambda^5\sqrt{1-\lambda^2}(\rho+i\eta),\nonumber\\
&V_{us}=\lambda\sqrt{1-A^2\lambda^6(\rho^2+\eta^2)}, ~~~ &V_{cs}&=\sqrt{1-\lambda^2}\sqrt{1-A^2\lambda^4}-A^2\lambda^6(\rho+i\eta),\label{Eq:ckmwolfenpar}
\end{align}
with $\lambda$, $A$, $\rho$ and $\eta$ are the real parameters. The latest results fitted by the UTfit collaboration are presented as following~\cite{Ref:utfit}
\begin{align}
&\lambda=0.2251\pm 0.0008, ~~ &A&=0.828\pm0.010,~~ &\rho&=0.164\pm0.009,~~ &\eta&=0.355\pm0.009.\label{Eq:ckmwolfenparval}
\end{align}
\subsection{The numerical results of the branching ratios}
\label{sec:numresbranchratios}
By substituting the values of the parameters listed in subsection \ref{sec:inputparameters} into Eqs.(\ref{Eq:decaywidthbct1}), (\ref{Eq:decaywidthbct2}), (\ref{Eq:decaywidthbct3}) and (\ref{Eq:decaywidthbct4}), we can obtain the numerical values of the branching ratios for the $B_{c}^{+}\rightarrow  B^{+} \pi^{\pm} e^{\mp} \nu_{e}$ and $B_{c}^{-}\rightarrow  B^{-} \pi^{\pm} e^{\mp} \nu_{e}$ decays, which are shown as following
\begin{small}
\begin{align}
&{\mathcal B}(B_{c}^{\pm}\rightarrow  B^{\pm} \pi^{\pm}  e^{\mp} \nu_{e})=\frac{{\mathcal B}(B_{c}^{+}\rightarrow  B^{+} \pi^{+}  e^{-} \bar{\nu}_{e})+ {\mathcal B}(B_{c}^{-}\rightarrow  B^{-} \pi^{-}  e^{+} \nu_{e}) }{2}=\left(5.17^{-0.52}_{+0.53}\right)\times 10^{-4},\label{Eq:numbtanchratio1}\\
&{\mathcal B}(B_{c}^{\pm}\rightarrow  B^{\pm} \pi^{\mp}  e^{\pm} \nu_{e})=\frac{{\mathcal B}(B_{c}^{+}\rightarrow  B^{+} \pi^{-}  e^{+} \nu_{e})+ {\mathcal B}(B_{c}^{-}\rightarrow  B^{-} \pi^{+}  e^{-} \bar{\nu}_{e}) }{2}=\left(5.13^{-0.51}_{+0.52}\right)\times 10^{-4},\label{Eq:numbtanchratio2}\\
&{\mathcal B}(B_{c}^{\pm}\rightarrow  B^{\pm} \pi^{+}  e^{-} \bar{\nu}_{e})=\frac{{\mathcal B}(B_{c}^{+}\rightarrow  B^{+} \pi^{+}  e^{-} \bar{\nu}_{e})+ {\mathcal B}(B_{c}^{-}\rightarrow  B^{-} \pi^{+}  e^{-} \bar{\nu}_{e}) }{2}=\left(5.13^{-0.51}_{+0.52}\right)\times 10^{-4},\label{Eq:numbtanchratio3}\\
&{\mathcal B}(B_{c}^{\pm}\rightarrow  B^{\pm} \pi^{-}  e^{+} \nu_{e})=\frac{{\mathcal B}(B_{c}^{+}\rightarrow  B^{+} \pi^{-}  e^{+} \nu_{e})+ {\mathcal B}(B_{c}^{-}\rightarrow  B^{-} \pi^{-}  e^{+} \nu_{e}) }{2}=\left(5.17^{-0.52}_{+0.53}\right)\times 10^{-4}.\label{Eq:numbtanchratio4}
\end{align}
\end{small}
By combining the above branching ratios with the branching ratios for the consequent decays of the $B^{\pm}$ mesons given by the Particle Data Group, we can calculate the results of the branching ratios ${\mathcal B}(B_{c}^{\pm}\rightarrow  B^{\pm} \pi^{\pm} e^{\mp} \nu_{e}\rightarrow f_{B^{\pm}} \pi^{\pm} e^{\mp} \nu_{e})$ (hereinafter for brevity collectively referred to as ${\mathcal B}(B_{c}^{\pm}\rightarrow  B^{\pm}\pi e \nu_{e}\rightarrow  f_{B^{\pm}} \pi e \nu_{e})$), which are equal to ${\mathcal B}(B_{c}^{\pm}\rightarrow  B^{\pm} \pi^{\pm}  e^{\mp} \nu_{e})\cdot {\mathcal B}(  B^{+} \rightarrow f_{B^{+}})$ and listed in Table~\ref{totbranchratio1}. We only consider the decay channels with the branching ratios are larger than $5.0\times 10^{-7}$ and the semileptonic decays of $B^{\pm}$ mesons, which are hopefully to be marginally observed by the current experiments.

According to these numerical results of ${\mathcal B}(B_{c}^{\pm}\rightarrow  B^{\pm}\pi e \nu_{e}\rightarrow  f_{B^{\pm}} \pi e \nu_{e})$ and the following formula~\cite{Zhou:2020bnm,Dai:1998hb,Fu:2011tn},
\begin{align}
(\epsilon_f N)_{\mathcal B}=\frac{9}{{\mathcal B}(B_{c}^{\pm}\rightarrow  B^{\pm} \pi e \nu_{e}\rightarrow f_{B^{\pm}} \pi e \nu_{e})},\label{Eq:numneedbrratio}
\end{align}
we can estimate that how many $B_{c}^{\pm}$ events-times-efficiency are needed to observe these decays with three standard deviations (3$\sigma$). In Eq.(\ref{Eq:numneedbrratio}), $\epsilon_f $ is the detecting efficiency of the final states, $N$ is the number of the $B_{c}^{\pm}$ meson needed to observe the decays with three standard deviations (3$\sigma$). With the numerical results of the branching ratios in Table~\ref{totbranchratio1}, we can obtain the numerical results of $(\epsilon_f N)_{\mathcal B}$ for the $B_{c}^{\pm}\rightarrow  B^{\pm} \pi^{\pm} e^{\mp} \nu_{e}\rightarrow f_{B^{\pm}} \pi^{\pm} e^{\mp} \nu_{e}$ channels, which are also listed in Table~\ref{totbranchratio1}.
\begin{table}[t]
\begin{center}
\caption{\label{totbranchratio1} \small The total branching fractions and the numerical results of $(\epsilon_f N)_{\mathcal B}$ for the $B_{c}^{\pm}\rightarrow  B^{\pm} \pi^{\pm} e^{\mp} \nu_{e}\rightarrow f_{B^{\pm}} \pi^{\pm} e^{\mp} \nu_{e}$ decays.}
\vspace{0.1cm}
\doublerulesep 0.8pt \tabcolsep 0.18in
\scriptsize
\begin{tabular}{c|c|c}
\hline
the $B^{+} \rightarrow f_{B^+}  $ decays & the branching ratio & $(\epsilon_f N)_{\mathcal B}$\\
\hline
$B^{+} \rightarrow\bar{D}^0 l^{+}\nu_l \rightarrow K^{+} \pi^{-} \pi^{0} l^{+}\nu_l $ & $\left(1.64^{-0.18}_{+0.19}\right)\times 10^{-6}$   & $\left(4.91\sim6.16\right)\times 10^{6}$ \\
\hline
$ B^{+} \rightarrow\bar{D}^0 l^{+}\nu_l \rightarrow K^{+} \pi^{-} \pi^{-}\pi^{+} l^{+}\nu_l $ & $\left(9.39^{-0.98}_{+1.00}\right)\times 10^{-7}$   & $\left(0.87\sim1.07\right)\times 10^{7}$ \\
\hline
$ B^{+} \rightarrow\bar{D}^0 l^{+}\nu_l \rightarrow K^{+} \pi^{-} \pi^{0}\pi^{0} l^{+}\nu_l $ & $\left(1.01\pm0.11\right)\times 10^{-6}$   & $\left(8.02\sim9.96\right)\times 10^{6}$ \\
\hline
$B^{+} \rightarrow\bar{D}^{*}(2007)^0 l^{+}\nu_l \rightarrow\bar{D}^{0}\pi^{0} l^{+}\nu_l \rightarrow K^{+} e^{-} \bar{\nu}_{e} \pi^{0} l^{+}\nu_l $ & $\left(6.56^{-0.71}_{+0.73}\right)\times 10^{-7}$   & $\left(1.23\sim1.54\right)\times 10^{7}$ \\
\hline
$B^{+} \rightarrow\bar{D}^{*}(2007)^0 l^{+}\nu_l \rightarrow\bar{D}^{0}\pi^{0} l^{+}\nu_l \rightarrow K^{+} \mu^{-} \bar{\nu}_{\mu} \pi^{0} l^{+}\nu_l $ & $\left(6.31^{-0.69}_{+0.70}\right)\times 10^{-7}$   & $\left(1.28\sim1.60\right)\times 10^{7}$ \\
\hline
$B^{+} \rightarrow\bar{D}^{*}(2007)^0 l^{+}\nu_l \rightarrow\bar{D}^{0}\pi^{0} l^{+}\nu_l \rightarrow K^{+} \pi^{-} \pi^{0} l^{+}\nu_l $ & $\left(7.30^{-0.79}_{+0.81}\right)\times 10^{-7}$   & $\left(1.11\sim1.38\right)\times 10^{7}$ \\
\hline
$B^{+} \rightarrow\bar{D}^{*}(2007)^0 l^{+}\nu_l \rightarrow\bar{D}^{0}\pi^{0} l^{+}\nu_l \rightarrow K^{+} \pi^{-}  \pi^{0} \pi^{0} l^{+}\nu_l $ & $\left(2.66\pm 0.31\right)\times 10^{-6}$   & $\left(3.02\sim3.82\right)\times 10^{6}$ \\
\hline
$B^{+} \rightarrow\bar{D}^{*}(2007)^0 l^{+}\nu_l \rightarrow\bar{D}^{0}\pi^{0} l^{+}\nu_l \rightarrow K^{+} \pi^{-} \pi^{-} \pi^{+}  \pi^{0} l^{+}\nu_l $ & $\left(1.52\pm 0.17\right)\times 10^{-6}$   & $\left(5.33\sim6.65\right)\times 10^{6}$ \\
\hline
$B^{+} \rightarrow\bar{D}^{*}(2007)^0 l^{+}\nu_l \rightarrow\bar{D}^{0}\pi^{0} l^{+}\nu_l \rightarrow K^{+} \pi^{-} \pi^{0} \pi^{0}  \pi^{0} l^{+}\nu_l $ & $\left(1.64^{-0.18}_{+0.19}\right)\times 10^{-6}$   & $\left(4.93\sim6.18\right)\times 10^{6}$ \\
\hline
$B^{+} \rightarrow\bar{D}^{*}(2007)^0 l^{+}\nu_l \rightarrow\bar{D}^{0}\pi^{0} l^{+}\nu_l \rightarrow K^{+} \pi^{-} \pi^{-} \pi^{+} \pi^{0}  \pi^{0} l^{+}\nu_l $ & $\left(7.95^{-1.14}_{+1.15}\right)\times 10^{-7}$   & $\left(0.99\sim1.32\right)\times 10^{7}$ \\
\hline
$B^{+} \rightarrow\bar{D}^{*}(2007)^0 l^{+}\nu_l \rightarrow\bar{D}^{0}\gamma l^{+}\nu_l \rightarrow K^{+} \pi^{-} \pi^{0}\gamma l^{+}\nu_l  $ & $\left(1.45\pm0.17\right)\times 10^{-6}$   & $\left(5.53\sim7.02\right)\times 10^{6}$ \\
\hline
$B^{+} \rightarrow\bar{D}^{*}(2007)^0 l^{+}\nu_l \rightarrow\bar{D}^{0}\gamma l^{+}\nu_l \rightarrow K^{+} \pi^{-} \pi^{-}\pi^{+}\gamma l^{+}\nu_l  $ & $\left(8.29^{-0.93}_{+0.94}\right)\times 10^{-7}$   & $\left(0.97\sim1.22\right)\times 10^{7}$ \\
\hline
$B^{+} \rightarrow\bar{D}^{*}(2007)^0 l^{+}\nu_l \rightarrow\bar{D}^{0}\gamma l^{+}\nu_l \rightarrow K^{+} \pi^{-} \pi^{0}\pi^{0}\gamma l^{+}\nu_l  $ & $\left(8.94^{-1.01}_{+1.03}\right)\times 10^{-7}$   & $\left(0.90\sim1.14\right)\times 10^{7}$ \\
\hline
\end{tabular}
\end{center}
\end{table}
Similarly, we can calculate the branching ratios and the numerical results of $(\epsilon_f N)_{\mathcal B}$ for the $B_{c}^{\pm}\rightarrow  B^{\pm} \pi^{\mp} e^{\pm} \nu_{e}\rightarrow f_{B^{\pm}} \pi^{\mp} e^{\pm} \nu_{e}$, $B_{c}^{\pm}\rightarrow  B^{\pm} \pi^{+} e^{-} \bar{\nu}_{e}\rightarrow f_{B^{\pm}} \pi^{+} e^{-} \bar{\nu}_{e}$ and $B_{c}^{\pm}\rightarrow  B^{\pm} \pi^{-} e^{+} \nu_{e}\rightarrow f_{B^{\pm}} \pi^{-} e^{+} \nu_{e}$ decays, which are listed in Table~\ref{totbranchratio2}, Table~\ref{totbranchratio3} and Table~\ref{totbranchratio4}, respectively. Here, we note that the branching ratios for the $B_{c}^{\pm}\rightarrow  B^{\pm} \pi^{\mp} e^{\pm} \nu_{e}\rightarrow f_{B^{\pm}} \pi^{\mp} e^{\pm} \nu_{e}$, $B_{c}^{\pm}\rightarrow  B^{\pm} \pi^{+} e^{-} \bar{\nu}_{e}\rightarrow f_{B^{\pm}} \pi^{+} e^{-} \bar{\nu}_{e}$ and $B_{c}^{\pm}\rightarrow  B^{\pm} \pi^{-} e^{+} \nu_{e}\rightarrow f_{B^{\pm}} \pi^{-} e^{+} \nu_{e}$ decays (hereinafter for brevity collectively referred to as ${\mathcal B}(B_{c}^{\pm}\rightarrow  B^{\pm}\pi e \nu_{e}\rightarrow  f_{B^{\pm}} \pi e \nu_{e})$) are equal to ${\mathcal B}(B_{c}^{\pm}\rightarrow  B^{\pm} \pi^{\mp}  e^{\pm} \nu_{e})\cdot {\mathcal B}(  B^{+} \rightarrow f_{B^{+}})$, ${\mathcal B}(B_{c}^{\pm}\rightarrow  B^{\pm} \pi^{+}  e^{-} \bar{\nu}_{e})\cdot {\mathcal B}(  B^{+} \rightarrow f_{B^{+}})$ and ${\mathcal B}(B_{c}^{\pm}\rightarrow  B^{\pm} \pi^{-}  e^{+} \nu_{e})\cdot {\mathcal B}(  B^{+} \rightarrow f_{B^{+}})$, respectively.
\begin{table}[t]
\begin{center}
\caption{\label{totbranchratio2} \small The total branching fractions and  the numerical results of $(\epsilon_f N)_{\mathcal B}$ for the $B_{c}^{\pm}\rightarrow  B^{\pm} \pi^{\mp} e^{\pm} \nu_{e}\rightarrow f_{B^{\pm}} \pi^{\mp} e^{\pm} \nu_{e}$ decays.}
\vspace{0.1cm}
\doublerulesep 0.8pt \tabcolsep 0.18in
\scriptsize
\begin{tabular}{c|c|c}
\hline
the $B^{+} \rightarrow f_{B^+}  $ decays & the branching ratio & $(\epsilon_f N)_{\mathcal B}$\\
\hline
$B^{+} \rightarrow\bar{D}^0 l^{+}\nu_l \rightarrow K^{+} \pi^{-} \pi^{0} l^{+}\nu_l $ & $\left(1.63^{-0.18}_{+0.19}\right)\times 10^{-6}$   & $\left(4.95\sim6.20\right)\times 10^{6}$ \\
\hline
$ B^{+} \rightarrow\bar{D}^0 l^{+}\nu_l \rightarrow K^{+} \pi^{-} \pi^{-}\pi^{+} l^{+}\nu_l $ & $\left(9.33^{-0.98}_{+1.00}\right)\times 10^{-7}$   & $\left(0.87\sim1.08\right)\times 10^{7}$ \\
\hline
$ B^{+} \rightarrow\bar{D}^0 l^{+}\nu_l \rightarrow K^{+} \pi^{-} \pi^{0}\pi^{0} l^{+}\nu_l $ & $\left(1.01\pm0.11\right)\times 10^{-6}$   & $\left(0.81\sim1.00\right)\times 10^{7}$ \\
\hline
$B^{+} \rightarrow\bar{D}^{*}(2007)^0 l^{+}\nu_l \rightarrow\bar{D}^{0}\pi^{0} l^{+}\nu_l \rightarrow K^{+} e^{-} \bar{\nu}_{e} \pi^{0} l^{+}\nu_l $ & $\left(6.52^{-0.71}_{+0.72}\right)\times 10^{-7}$   & $\left(1.24\sim1.55\right)\times 10^{7}$ \\
\hline
$B^{+} \rightarrow\bar{D}^{*}(2007)^0 l^{+}\nu_l \rightarrow\bar{D}^{0}\pi^{0} l^{+}\nu_l \rightarrow K^{+} \mu^{-} \bar{\nu}_{\mu} \pi^{0} l^{+}\nu_l $ & $\left(6.26^{-0.68}_{+0.70}\right)\times 10^{-7}$   & $\left(1.29\sim1.61\right)\times 10^{7}$ \\
\hline
$B^{+} \rightarrow\bar{D}^{*}(2007)^0 l^{+}\nu_l \rightarrow\bar{D}^{0}\pi^{0} l^{+}\nu_l \rightarrow K^{+} \pi^{-} \pi^{0} l^{+}\nu_l $ & $\left(7.25^{-0.79}_{+0.80}\right)\times 10^{-7}$   & $\left(1.12\sim1.39\right)\times 10^{7}$ \\
\hline
$B^{+} \rightarrow\bar{D}^{*}(2007)^0 l^{+}\nu_l \rightarrow\bar{D}^{0}\pi^{0} l^{+}\nu_l \rightarrow K^{+} \pi^{-}  \pi^{0} \pi^{0} l^{+}\nu_l $ & $\left(2.65\pm 0.31\right)\times 10^{-6}$   & $\left(3.04\sim3.85\right)\times 10^{6}$ \\
\hline
$B^{+} \rightarrow\bar{D}^{*}(2007)^0 l^{+}\nu_l \rightarrow\bar{D}^{0}\pi^{0} l^{+}\nu_l \rightarrow K^{+} \pi^{-} \pi^{-} \pi^{+}  \pi^{0} l^{+}\nu_l $ & $\left(1.51\pm 0.17\right)\times 10^{-6}$   & $\left(5.36\sim6.70\right)\times 10^{6}$ \\
\hline
$B^{+} \rightarrow\bar{D}^{*}(2007)^0 l^{+}\nu_l \rightarrow\bar{D}^{0}\pi^{0} l^{+}\nu_l \rightarrow K^{+} \pi^{-} \pi^{0} \pi^{0}  \pi^{0} l^{+}\nu_l $ & $\left(1.63\pm0.18\right)\times 10^{-6}$   & $\left(4.97\sim6.22\right)\times 10^{6}$ \\
\hline
$B^{+} \rightarrow\bar{D}^{*}(2007)^0 l^{+}\nu_l \rightarrow\bar{D}^{0}\pi^{0} l^{+}\nu_l \rightarrow K^{+} \pi^{-} \pi^{-} \pi^{+} \pi^{0}  \pi^{0} l^{+}\nu_l $ & $\left(7.90^{-1.13}_{+1.14}\right)\times 10^{-7}$   & $\left(1.00\sim1.33\right)\times 10^{7}$ \\
\hline
$B^{+} \rightarrow\bar{D}^{*}(2007)^0 l^{+}\nu_l \rightarrow\bar{D}^{0}\gamma l^{+}\nu_l \rightarrow K^{+} \pi^{-} \pi^{0}\gamma l^{+}\nu_l  $ & $\left(1.44\pm0.17\right)\times 10^{-6}$   & $\left(5.57\sim7.07\right)\times 10^{6}$ \\
\hline
$B^{+} \rightarrow\bar{D}^{*}(2007)^0 l^{+}\nu_l \rightarrow\bar{D}^{0}\gamma l^{+}\nu_l \rightarrow K^{+} \pi^{-} \pi^{-}\pi^{+}\gamma l^{+}\nu_l  $ & $\left(8.24^{-0.92}_{+0.94}\right)\times 10^{-7}$   & $\left(0.98\sim1.23\right)\times 10^{7}$ \\
\hline
$B^{+} \rightarrow\bar{D}^{*}(2007)^0 l^{+}\nu_l \rightarrow\bar{D}^{0}\gamma l^{+}\nu_l \rightarrow K^{+} \pi^{-} \pi^{0}\pi^{0}\gamma l^{+}\nu_l  $ & $\left(8.88^{-1.01}_{+1.03}\right)\times 10^{-7}$   & $\left(0.91\sim1.14\right)\times 10^{7}$ \\
\hline
\end{tabular}
\end{center}
\end{table}
\begin{table}[t]
\begin{center}
\caption{\label{totbranchratio3} \small The total branching fractions and  the numerical results of $(\epsilon_f N)_{\mathcal B}$ for the $B_{c}^{\pm}\rightarrow  B^{\pm} \pi^{+} e^{-} \bar{\nu}_{e}\rightarrow f_{B^{\pm}} \pi^{+} e^{-} \bar{\nu}_{e}$  decays.}
\vspace{0.1cm}
\doublerulesep 0.8pt \tabcolsep 0.18in
\scriptsize
\begin{tabular}{c|c|c}
\hline
the $B^{+} \rightarrow f_{B^+}  $ decays & the branching ratio & $(\epsilon_f N)_{\mathcal B}$\\
\hline
$B^{+} \rightarrow\bar{D}^0 l^{+}\nu_l \rightarrow K^{+} \pi^{-} \pi^{0} l^{+}\nu_l $ & $\left(1.63^{-0.18}_{+0.19}\right)\times 10^{-6}$   & $\left(4.95\sim6.20\right)\times 10^{6}$ \\
\hline
$ B^{+} \rightarrow\bar{D}^0 l^{+}\nu_l \rightarrow K^{+} \pi^{-} \pi^{-}\pi^{+} l^{+}\nu_l $ & $\left(9.33^{-0.98}_{+1.00}\right)\times 10^{-7}$   & $\left(0.87\sim1.08\right)\times 10^{7}$ \\
\hline
$ B^{+} \rightarrow\bar{D}^0 l^{+}\nu_l \rightarrow K^{+} \pi^{-} \pi^{0}\pi^{0} l^{+}\nu_l $ & $\left(1.01\pm0.11\right)\times 10^{-6}$   & $\left(0.81\sim1.00\right)\times 10^{7}$ \\
\hline
$B^{+} \rightarrow\bar{D}^{*}(2007)^0 l^{+}\nu_l \rightarrow\bar{D}^{0}\pi^{0} l^{+}\nu_l \rightarrow K^{+} e^{-} \bar{\nu}_{e} \pi^{0} l^{+}\nu_l $ & $\left(6.52^{-0.71}_{+0.72}\right)\times 10^{-7}$   & $\left(1.24\sim1.55\right)\times 10^{7}$ \\
\hline
$B^{+} \rightarrow\bar{D}^{*}(2007)^0 l^{+}\nu_l \rightarrow\bar{D}^{0}\pi^{0} l^{+}\nu_l \rightarrow K^{+} \mu^{-} \bar{\nu}_{\mu} \pi^{0} l^{+}\nu_l $ & $\left(6.26^{-0.68}_{+0.70}\right)\times 10^{-7}$   & $\left(1.29\sim1.61\right)\times 10^{7}$ \\
\hline
$B^{+} \rightarrow\bar{D}^{*}(2007)^0 l^{+}\nu_l \rightarrow\bar{D}^{0}\pi^{0} l^{+}\nu_l \rightarrow K^{+} \pi^{-} \pi^{0} l^{+}\nu_l $ & $\left(7.25^{-0.79}_{+0.80}\right)\times 10^{-7}$   & $\left(1.12\sim1.39\right)\times 10^{7}$ \\
\hline
$B^{+} \rightarrow\bar{D}^{*}(2007)^0 l^{+}\nu_l \rightarrow\bar{D}^{0}\pi^{0} l^{+}\nu_l \rightarrow K^{+} \pi^{-}  \pi^{0} \pi^{0} l^{+}\nu_l $ & $\left(2.65\pm 0.31\right)\times 10^{-6}$   & $\left(3.04\sim3.85\right)\times 10^{6}$ \\
\hline
$B^{+} \rightarrow\bar{D}^{*}(2007)^0 l^{+}\nu_l \rightarrow\bar{D}^{0}\pi^{0} l^{+}\nu_l \rightarrow K^{+} \pi^{-} \pi^{-} \pi^{+}  \pi^{0} l^{+}\nu_l $ & $\left(1.51\pm 0.17\right)\times 10^{-6}$   & $\left(5.36\sim6.70\right)\times 10^{6}$ \\
\hline
$B^{+} \rightarrow\bar{D}^{*}(2007)^0 l^{+}\nu_l \rightarrow\bar{D}^{0}\pi^{0} l^{+}\nu_l \rightarrow K^{+} \pi^{-} \pi^{0} \pi^{0}  \pi^{0} l^{+}\nu_l $ & $\left(1.63\pm0.18\right)\times 10^{-6}$   & $\left(4.97\sim6.22\right)\times 10^{6}$ \\
\hline
$B^{+} \rightarrow\bar{D}^{*}(2007)^0 l^{+}\nu_l \rightarrow\bar{D}^{0}\pi^{0} l^{+}\nu_l \rightarrow K^{+} \pi^{-} \pi^{-} \pi^{+} \pi^{0}  \pi^{0} l^{+}\nu_l $ & $\left(7.90^{-1.13}_{+1.14}\right)\times 10^{-7}$   & $\left(1.00\sim1.33\right)\times 10^{7}$ \\
\hline
$B^{+} \rightarrow\bar{D}^{*}(2007)^0 l^{+}\nu_l \rightarrow\bar{D}^{0}\gamma l^{+}\nu_l \rightarrow K^{+} \pi^{-} \pi^{0}\gamma l^{+}\nu_l  $ & $\left(1.44\pm0.17\right)\times 10^{-6}$   & $\left(5.57\sim7.07\right)\times 10^{6}$ \\
\hline
$B^{+} \rightarrow\bar{D}^{*}(2007)^0 l^{+}\nu_l \rightarrow\bar{D}^{0}\gamma l^{+}\nu_l \rightarrow K^{+} \pi^{-} \pi^{-}\pi^{+}\gamma l^{+}\nu_l  $ & $\left(8.24^{-0.92}_{+0.94}\right)\times 10^{-7}$   & $\left(0.98\sim1.23\right)\times 10^{7}$ \\
\hline
$B^{+} \rightarrow\bar{D}^{*}(2007)^0 l^{+}\nu_l \rightarrow\bar{D}^{0}\gamma l^{+}\nu_l \rightarrow K^{+} \pi^{-} \pi^{0}\pi^{0}\gamma l^{+}\nu_l  $ & $\left(8.88^{-1.01}_{+1.03}\right)\times 10^{-7}$   & $\left(0.91\sim1.14\right)\times 10^{7}$ \\
\hline
\end{tabular}
\end{center}
\end{table}
\begin{table}[t]
\begin{center}
\caption{\label{totbranchratio4} \small The total branching fractions and  the numerical results of $(\epsilon_f N)_{\mathcal B}$ for the $B_{c}^{\pm}\rightarrow  B^{\pm} \pi^{-} e^{+} \nu_{e}\rightarrow f_{B^{\pm}} \pi^{-} e^{+} \nu_{e}$ decays.}
\vspace{0.1cm}
\doublerulesep 0.8pt \tabcolsep 0.18in
\scriptsize
\begin{tabular}{c|c|c}
\hline
the $B^{+} \rightarrow f_{B^+}  $ decays & the branching ratio & $(\epsilon_f N)_{\mathcal B}$\\
\hline
$B^{+} \rightarrow\bar{D}^0 l^{+}\nu_l \rightarrow K^{+} \pi^{-} \pi^{0} l^{+}\nu_l $ & $\left(1.64^{-0.18}_{+0.19}\right)\times 10^{-6}$   & $\left(4.91\sim6.16\right)\times 10^{6}$ \\
\hline
$ B^{+} \rightarrow\bar{D}^0 l^{+}\nu_l \rightarrow K^{+} \pi^{-} \pi^{-}\pi^{+} l^{+}\nu_l $ & $\left(9.39^{-0.98}_{+1.00}\right)\times 10^{-7}$   & $\left(0.87\sim1.07\right)\times 10^{7}$ \\
\hline
$ B^{+} \rightarrow\bar{D}^0 l^{+}\nu_l \rightarrow K^{+} \pi^{-} \pi^{0}\pi^{0} l^{+}\nu_l $ & $\left(1.01\pm0.11\right)\times 10^{-6}$   & $\left(8.02\sim9.
96\right)\times 10^{6}$ \\
\hline
$B^{+} \rightarrow\bar{D}^{*}(2007)^0 l^{+}\nu_l \rightarrow\bar{D}^{0}\pi^{0} l^{+}\nu_l \rightarrow K^{+} e^{-} \bar{\nu}_{e} \pi^{0} l^{+}\nu_l $ & $\left(6.56^{-0.71}_{+0.73}\right)\times 10^{-7}$   & $\left(1.23\sim1.54\right)\times 10^{7}$ \\
\hline
$B^{+} \rightarrow\bar{D}^{*}(2007)^0 l^{+}\nu_l \rightarrow\bar{D}^{0}\pi^{0} l^{+}\nu_l \rightarrow K^{+} \mu^{-} \bar{\nu}_{\mu} \pi^{0} l^{+}\nu_l $ & $\left(6.31^{-0.69}_{+0.70}\right)\times 10^{-7}$   & $\left(1.28\sim1.60\right)\times 10^{7}$ \\
\hline
$B^{+} \rightarrow\bar{D}^{*}(2007)^0 l^{+}\nu_l \rightarrow\bar{D}^{0}\pi^{0} l^{+}\nu_l \rightarrow K^{+} \pi^{-} \pi^{0} l^{+}\nu_l $ & $\left(7.30^{-0.79}_{+0.81}\right)\times 10^{-7}$   & $\left(1.11\sim1.38\right)\times 10^{7}$ \\
\hline
$B^{+} \rightarrow\bar{D}^{*}(2007)^0 l^{+}\nu_l \rightarrow\bar{D}^{0}\pi^{0} l^{+}\nu_l \rightarrow K^{+} \pi^{-}  \pi^{0} \pi^{0} l^{+}\nu_l $ & $\left(2.66\pm 0.31\right)\times 10^{-6}$   & $\left(3.02\sim3.82\right)\times 10^{6}$ \\
\hline
$B^{+} \rightarrow\bar{D}^{*}(2007)^0 l^{+}\nu_l \rightarrow\bar{D}^{0}\pi^{0} l^{+}\nu_l \rightarrow K^{+} \pi^{-} \pi^{-} \pi^{+}  \pi^{0} l^{+}\nu_l $ & $\left(1.52\pm 0.17\right)\times 10^{-6}$   & $\left(5.33\sim6.65\right)\times 10^{6}$ \\
\hline
$B^{+} \rightarrow\bar{D}^{*}(2007)^0 l^{+}\nu_l \rightarrow\bar{D}^{0}\pi^{0} l^{+}\nu_l \rightarrow K^{+} \pi^{-} \pi^{0} \pi^{0}  \pi^{0} l^{+}\nu_l $ & $\left(1.64^{-0.18}_{+0.19}\right)\times 10^{-6}$   & $\left(4.93\sim6.18\right)\times 10^{6}$ \\
\hline
$B^{+} \rightarrow\bar{D}^{*}(2007)^0 l^{+}\nu_l \rightarrow\bar{D}^{0}\pi^{0} l^{+}\nu_l \rightarrow K^{+} \pi^{-} \pi^{-} \pi^{+} \pi^{0}  \pi^{0} l^{+}\nu_l $ & $\left(7.95^{-1.14}_{+1.15}\right)\times 10^{-7}$   & $\left(0.99\sim1.32\right)\times 10^{7}$ \\
\hline
$B^{+} \rightarrow\bar{D}^{*}(2007)^0 l^{+}\nu_l \rightarrow\bar{D}^{0}\gamma l^{+}\nu_l \rightarrow K^{+} \pi^{-} \pi^{0}\gamma l^{+}\nu_l  $ & $\left(1.45\pm0.17\right)\times 10^{-6}$   & $\left(5.53\sim7.02\right)\times 10^{6}$ \\
\hline
$B^{+} \rightarrow\bar{D}^{*}(2007)^0 l^{+}\nu_l \rightarrow\bar{D}^{0}\gamma l^{+}\nu_l \rightarrow K^{+} \pi^{-} \pi^{-}\pi^{+}\gamma l^{+}\nu_l  $ & $\left(8.29^{-0.93}_{+0.94}\right)\times 10^{-7}$   & $\left(0.97\sim1.22\right)\times 10^{7}$ \\
\hline
$B^{+} \rightarrow\bar{D}^{*}(2007)^0 l^{+}\nu_l \rightarrow\bar{D}^{0}\gamma l^{+}\nu_l \rightarrow K^{+} \pi^{-} \pi^{0}\pi^{0}\gamma l^{+}\nu_l  $ & $\left(8.94^{-1.01}_{+1.03}\right)\times 10^{-7}$   & $\left(0.90\sim1.14\right)\times 10^{7}$ \\
\hline
\end{tabular}
\end{center}
\end{table}
\subsection{The numerical results of the CP asymmetries}
\label{sec:numrescpasymmetry}
Now, we proceed to calculate the numerical results of the CP asymmetries in $B_{c}^{\pm}\rightarrow  B^{\pm} \pi^{\pm} e^{\mp} \nu_{e}\rightarrow f_{B^{\pm}} \pi^{\pm} e^{\mp} \nu_{e}$ and $B_{c}^{\pm}\rightarrow  B^{\pm} \pi^{\mp} e^{\pm} \nu_{e}\rightarrow f_{B^{\pm}} \pi^{\mp} e^{\pm} \nu_{e}$ decays. By substituting the values of the parameters in subsection \ref{sec:inputparameters} into Eqs.(\ref{Eq:cpasymmacppm31})-(\ref{Eq:cpasymmacppm34}) and Eqs.(\ref{Eq:cpasymmacpmp31})-(\ref{Eq:cpasymmacpmp34}), we can obtain the numerical results of the CP asymmetries in $B_{c}^{\pm}\rightarrow  B^{\pm} \pi^{\pm} e^{\mp} \nu_{e}\rightarrow f_{B^{\pm}} \pi^{\pm} e^{\mp} \nu_{e}$ and $B_{c}^{\pm}\rightarrow  B^{\pm} \pi^{\mp} e^{\pm} \nu_{e}\rightarrow f_{B^{\pm}} \pi^{\mp} e^{\pm} \nu_{e}$ decays, which are shown in Table~\ref{cpasymmetrunum}.
\begin{table}[t]
\begin{center}
\caption{\label{cpasymmetrunum} \small The numerical results of the CP asymmetries in $B_{c}^{\pm}\rightarrow  B^{\pm} \pi^{\pm} e^{\mp} \nu_{e}\rightarrow f_{B^{\pm}} \pi^{\pm} e^{\mp} \nu_{e}$ and $B_{c}^{\pm}\rightarrow  B^{\pm} \pi^{\mp} e^{\pm} \nu_{e}\rightarrow f_{B^{\pm}} \pi^{\mp} e^{\pm} \nu_{e}$ decays.}
\vspace{0.1cm}
\doublerulesep 0.8pt \tabcolsep 0.18in
\scriptsize
\begin{tabular}{c|c}
\hline
${\mathcal A}_{CP}^{pm}=\left(-3.36\pm0.05\right)\times 10^{-4}$ & ${\mathcal A}_{CP}^{mp}=\left(6.30\pm0.08\right)\times 10^{-3}$       \\
${\mathcal A}_{CP,mix}^{pm}=\left(-1.69\pm0.03\right)\times 10^{-5}$ & ${\mathcal A}_{CP,mix}^{mp}=\left(5.98\pm0.07\right)\times 10^{-3}$\\
${\mathcal A}_{CP,dir}^{pm}=\left(-4.06\pm0.17\right)\times 10^{-7}$ & ${\mathcal A}_{CP,dir}^{mp}=\left(2.20\pm0.28\right)\times 10^{-8}$\\
${\mathcal A}_{CP,int}^{pm}=\left(-3.18\pm0.04\right)\times 10^{-4}$ & ${\mathcal A}_{CP,int}^{mp}=\left(3.21\pm0.04\right)\times 10^{-4}$\\
\hline
\end{tabular}
\end{center}
\end{table}
From these numerical values, we can obtain the following points:
\begin{enumerate}
\item Using the numerical results of the parameters in subsection \ref{sec:inputparameters}: $\Gamma_{S}/ \Gamma_{L}\approx 571$, $\Delta m / \Gamma_{L}\approx 271$, $\Gamma / \Gamma_{L}\approx 286$ and Eqs.(\ref{Eq:gpmtintgrate1})-(\ref{Eq:gpmtintgrate3}), we can obtain
    \begin{align}
    &~~~~~~~~~~\int_{0}^{\infty}\left(g_{-}(t)\right)^{\ast} g_{+}(t) dt\approx 1.94\times 10^{16}+I\cdot 6.78\times 10^{13},\label{Eq:numimregmgptabs}\\
    &\int_{0}^{\infty}\left|g_{-}(t)\right|^2 dt \approx  1.94\times 10^{16}, ~~~~~~\int_{0}^{\infty}\left|g_{+}(t)\right|^2 dt\approx 1.95\times 10^{16},\label{Eq:numimregmpabssq}
    \end{align}
    so $r_g\approx 0.99 $, $r_{sf}\approx 0.99 $ and $\delta\approx -3.50\times 10^{-3}\text{rad}$, which can be derived from Eq.(\ref{Eq:definrwfrsfrg}).
\item Using the numerical results of the CKM matrix elements in subsection \ref{sec:inputparameters} and Eq.(\ref{Eq:definrwfrsfrg}), we can obtain
     \begin{align}
    &r_{wf}\approx 0.053,~~~~~~~\phi\approx-6.42\times 10^{-4}\text{rad}.\label{Eq:numimregmrwfphi}
    \end{align}
\item From Eq.(\ref{Eq:cpasymmacppmstrph}) and Eq.(\ref{Eq:cpasymmacpmpstrph}), we can see that the strong phase differences $\delta_{S}^{pm}$ and  $\delta_{S}^{mp}$ suffer from  the $Im(\epsilon)$ and $\delta$ suppressions, their numerical values are very small
    \begin{align}
    &\delta_{S}^{pm}\approx -6.64\times 10^{-3}\text{rad},~~~~~~~\delta_{S}^{mp}\approx-3.55\times 10^{-4}\text{rad}.\label{Eq:numstrongphasepmmp}
    \end{align}
\item  As can be seen from Eq.(\ref{Eq:cpasymmacppm33}) and Eq.(\ref{Eq:cpasymmacpmp33}), the direct CP asymmetry ${\mathcal A}_{CP,dir}^{pm}$ (${\mathcal A}_{CP,dir}^{mp}$) suffers from the $r_{wf}$, $\sin\phi$ and $\sin\delta_{S}^{pm}$ ($\sin\delta_{S}^{mp}$) suppressions, thus its numerical value is of the order of $10^{-7}$ ($10^{-8}$), which can be neglected safely.
\item As shown in Eq.(\ref{Eq:cpasymmacppm32}) and Eq.(\ref{Eq:cpasymmacppm35}), the new CP violation effect, i.e. the CP violation in the interference, ${\mathcal A}_{CP,int}^{pm}$ in $B_{c}^{\pm}\rightarrow  B^{\pm} \pi^{\pm} e^{\mp} \nu_{e}\rightarrow f_{B^{\pm}} \pi^{\pm} e^{\mp} \nu_{e}$ decays contains the term proportional to $\cos\left(\phi-\delta\right)$, so it merely suffers from the $r_{wf}$ and $Re\epsilon$ suppression, while the indirect CP violation in kaon mixing ${\mathcal A}_{CP,mix}^{pm}$ in $B_{c}^{\pm}\rightarrow  B^{\pm} \pi^{\pm} e^{\mp} \nu_{e}\rightarrow f_{B^{\pm}} \pi^{\pm} e^{\mp} \nu_{e}$ decays suffers from both $r_{wf}^2$ and $\left|q/p\right|^2-\left|p/q\right|^2$ suppressions, so the new CP violation effect ${\mathcal A}_{CP,int}^{pm}$ is dominant in the CP asymmetry ${\mathcal A}_{CP}^{pm}$ in $B_{c}^{\pm}\rightarrow  B^{\pm} \pi^{\pm} e^{\mp} \nu_{e}\rightarrow f_{B^{\pm}} \pi^{\pm} e^{\mp} \nu_{e}$ decays, the indirect CP violation in kaon mixing ${\mathcal A}_{CP,mix}^{pm}$ only accounts for about $5\%$ of the CP asymmetry ${\mathcal A}_{CP}^{pm}$ in $B_{c}^{\pm}\rightarrow  B^{\pm} \pi^{\pm} e^{\mp} \nu_{e}\rightarrow f_{B^{\pm}} \pi^{\pm} e^{\mp} \nu_{e}$ decays.
\item As for the CP asymmetry ${\mathcal A}_{CP}^{mp}$ in $B_{c}^{\pm}\rightarrow  B^{\pm} \pi^{\mp} e^{\pm} \nu_{e}\rightarrow f_{B^{\pm}} \pi^{\mp} e^{\pm} \nu_{e}$ decays, the indirect CP violation in kaon mixing ${\mathcal A}_{CP,mix}^{mp}$ merely suffers from the $\left|p/q\right|^2-\left|q/p\right|^2$ suppression, while the new CP violation ${\mathcal A}_{CP,int}^{mp}$ suffers from the $r_{wf}$ and $Re\epsilon$ suppression, so the indirect CP violation in kaon mixing ${\mathcal A}_{CP,mix}^{mp}$ is dominant in the CP asymmetry ${\mathcal A}_{CP}^{mp}$ in $B_{c}^{\pm}\rightarrow  B^{\pm} \pi^{\mp} e^{\pm} \nu_{e}\rightarrow f_{B^{\pm}} \pi^{\mp} e^{\pm} \nu_{e}$ decays, the new CP violation effect ${\mathcal A}_{CP,int}^{mp}$ only accounts for about $5\%$ of the CP asymmetry ${\mathcal A}_{CP}^{mp}$ in $B_{c}^{\pm}\rightarrow  B^{\pm} \pi^{\mp} e^{\pm} \nu_{e}\rightarrow f_{B^{\pm}} \pi^{\mp} e^{\pm} \nu_{e}$ decays, all these conclusions can be obtained from Eq.(\ref{Eq:cpasymmacpmp32}) and Eq.(\ref{Eq:cpasymmacpmp35}).
\end{enumerate}
In a word, the new CP violation  effect ${\mathcal A}_{CP,int}^{pm}$ dominates the CP asymmetry ${\mathcal A}_{CP}^{pm}$ in $B_{c}^{\pm}\rightarrow  B^{\pm} \pi^{\pm} e^{\mp} \nu_{e}\rightarrow f_{B^{\pm}} \pi^{\pm} e^{\mp} \nu_{e}$ decays, the indirect CP violation in kaon mixing ${\mathcal A}_{CP,mix}^{mp}$ dominates the CP asymmetry ${\mathcal A}_{CP}^{mp}$ in $B_{c}^{\pm}\rightarrow  B^{\pm} \pi^{\mp} e^{\pm} \nu_{e}\rightarrow f_{B^{\pm}} \pi^{\mp} e^{\pm} \nu_{e}$ decays, so the CP asymmetry ${\mathcal A}_{CP}^{pm}$ in $B_{c}^{\pm}\rightarrow  B^{\pm} \pi^{\pm} e^{\mp} \nu_{e}\rightarrow f_{B^{\pm}} \pi^{\pm} e^{\mp} \nu_{e}$ decays provides an ideal place to study the new CP violation effect.

According to the numerical results of the CP asymmetries in Table~\ref{cpasymmetrunum}, we can estimate that how many $B_{c}$ events-times-efficiency are needed to establish the CP asymmetries to three standard deviations (3$\sigma$). When the CP violation ${\mathcal A}_{CP}^{pm}$ in $B_{c}^{\pm}\rightarrow  B^{\pm} \pi^{\pm} e^{\mp} \nu_{e}\rightarrow f_{B^{\pm}} \pi^{\pm} e^{\mp} \nu_{e}$ decays is observed at three standard deviations (3$\sigma$) level, the number of $B_{c}$ events-times-efficiency needed reads
\begin{align}
(\epsilon_f N)_{CP}^{pm}=&\frac{9}{\left[{\mathcal B}(B_c^+ \rightarrow B^{+} \pi^{+} e^{-} \bar{\nu}_{e}\rightarrow f_{B^+} \pi^{+} e^{-} \bar{\nu}_{e})+{\mathcal B}(B_c^- \rightarrow B^{-} \pi^{-} e^{+} \nu_{e}\rightarrow f_{B^-} \pi^{-} e^{+} \nu_{e})\right]\cdot \left| {\mathcal A}_{CP}^{pm}\right|}\nonumber\\
\approx &\frac{9}{2\cdot {\mathcal B}(B_{c}^{\pm}\rightarrow  B^{\pm} \pi^{\pm} e^{\mp} \nu_{e})\cdot {\mathcal B}(B^+\rightarrow f_{B^+})\cdot \left|{\mathcal A}_{CP}^{pm}\right|}.\label{Eq:numneedcppm}
\end{align}
By substituting the numerical results of ${\mathcal B}(B_{c}^{\pm}\rightarrow  B^{\pm} \pi^{\pm} e^{\mp} \nu_{e})\cdot {\mathcal B}(B^+\rightarrow f_{B^+})$ in Table~\ref{totbranchratio1} and that of ${\mathcal A}_{CP}^{pm}$ in  Table~\ref{cpasymmetrunum} into Eq.(\ref{Eq:numneedcppm}), we can obtain the numerical results of $(\epsilon_f N)_{CP}^{pm}$, which are listed in Table~\ref{tabnmneedcpasympm}.

Similarly, the numbers of $B_{c}$ events-times-efficiency needed to observe the CP violation ${\mathcal A}_{CP}^{mp}$ in $B_{c}^{\pm}\rightarrow  B^{\pm} \pi^{\mp} e^{\pm} \nu_{e}\rightarrow f_{B^{\pm}} \pi^{\mp} e^{\pm} \nu_{e}$ decays at a significance of 3$\sigma$ reads
\begin{align}
(\epsilon_f N)_{CP}^{mp}=&\frac{9}{\left[{\mathcal B}(B_c^+ \rightarrow B^{+} \pi^{-} e^{+} \nu_{e}\rightarrow f_{B^+} \pi^{-} e^{+} \nu_{e})+{\mathcal B}(B_c^- \rightarrow B^{-} \pi^{+} e^{-} \bar{\nu}_{e}\rightarrow f_{B^-} \pi^{+} e^{-} \bar{\nu}_{e})\right]\cdot \left| {\mathcal A}_{CP}^{mp}\right|}\nonumber\\
\approx &\frac{9}{2\cdot {\mathcal B}(B_{c}^{\pm}\rightarrow  B^{\pm} \pi^{\mp} e^{\pm} \nu_{e})\cdot {\mathcal B}(B^+\rightarrow f_{B^+})\cdot \left|{\mathcal A}_{CP}^{mp}\right|}.\label{Eq:numneedcpmp}
\end{align}
Using the numerical results of ${\mathcal B}(B_{c}^{\pm}\rightarrow B^{\pm} \pi^{\mp} e^{\pm} \nu_{e})\cdot {\mathcal B}(B^+\rightarrow f_{B^+})$ in Table~\ref{totbranchratio2}, the numerical results of ${\mathcal A}_{CP}^{mp}$ in  Table~\ref{cpasymmetrunum} and  Eq.(\ref{Eq:numneedcpmp}), we can obtain the numerical results of $(\epsilon_f N)_{CP}^{mp}$, which are also listed in Table~\ref{tabnmneedcpasympm}.
\begin{table}[t]
\begin{center}
\caption{\label{tabnmneedcpasympm} \small The numerical results of $(\epsilon_f N)_{CP}^{pm}$ and $(\epsilon_f N)_{CP}^{mp}$.}
\vspace{0.1cm}
\doublerulesep 0.8pt \tabcolsep 0.18in
\scriptsize
\begin{tabular}{c|c|c}
\hline
the $B^{+} \rightarrow f_{B^+}  $ decays & $(\epsilon_f N)_{CP}^{pm}$ & $(\epsilon_f N)_{CP}^{mp}$\\
\hline
$B^{+} \rightarrow\bar{D}^0 l^{+}\nu_l \rightarrow K^{+} \pi^{-} \pi^{0} l^{+}\nu_l $ & $\left(7.33\sim9.21\right)\times 10^{9}$   & $\left(3.92\sim4.92\right)\times 10^{8}$ \\
\hline
$ B^{+} \rightarrow\bar{D}^0 l^{+}\nu_l \rightarrow K^{+} \pi^{-} \pi^{-}\pi^{+} l^{+}\nu_l $ & $\left(1.29\sim1.60\right)\times 10^{10}$   & $\left(6.91\sim8.56\right)\times 10^{8}$ \\
\hline
$ B^{+} \rightarrow\bar{D}^0 l^{+}\nu_l \rightarrow K^{+} \pi^{-} \pi^{0}\pi^{0} l^{+}\nu_l $ & $\left(1.20\sim1.49\right)\times 10^{10}$   & $\left(6.40\sim7.
95\right)\times 10^{8}$ \\
\hline
$B^{+} \rightarrow\bar{D}^{*}(2007)^0 l^{+}\nu_l \rightarrow\bar{D}^{0}\pi^{0} l^{+}\nu_l \rightarrow K^{+} e^{-} \bar{\nu}_{e} \pi^{0} l^{+}\nu_l $ & $\left(1.84\sim2.30\right)\times 10^{10}$    & $\left(0.99\sim1.23\right)\times 10^{9}$ \\
\hline
$B^{+} \rightarrow\bar{D}^{*}(2007)^0 l^{+}\nu_l \rightarrow\bar{D}^{0}\pi^{0} l^{+}\nu_l \rightarrow K^{+} \mu^{-} \bar{\nu}_{\mu} \pi^{0} l^{+}\nu_l $ & $\left(1.92\sim2.39\right)\times 10^{10}$    & $\left(1.02\sim1.28\right)\times 10^{9}$ \\
\hline
$B^{+} \rightarrow\bar{D}^{*}(2007)^0 l^{+}\nu_l \rightarrow\bar{D}^{0}\pi^{0} l^{+}\nu_l \rightarrow K^{+} \pi^{-} \pi^{0} l^{+}\nu_l $ & $\left(1.66\sim2.07\right)\times 10^{10}$    & $\left(0.89\sim1.11\right)\times 10^{9}$ \\
\hline
$B^{+} \rightarrow\bar{D}^{*}(2007)^0 l^{+}\nu_l \rightarrow\bar{D}^{0}\pi^{0} l^{+}\nu_l \rightarrow K^{+} \pi^{-}  \pi^{0} \pi^{0} l^{+}\nu_l $ & $\left(4.51\sim5.72\right)\times 10^{9}$    & $\left(2.41\sim3.06\right)\times 10^{8}$ \\
\hline
$B^{+} \rightarrow\bar{D}^{*}(2007)^0 l^{+}\nu_l \rightarrow\bar{D}^{0}\pi^{0} l^{+}\nu_l \rightarrow K^{+} \pi^{-} \pi^{-} \pi^{+}  \pi^{0} l^{+}\nu_l $ & $\left(7.94\sim9.94\right)\times 10^{9}$    & $\left(4.25\sim5.31\right)\times 10^{8}$ \\
\hline
$B^{+} \rightarrow\bar{D}^{*}(2007)^0 l^{+}\nu_l \rightarrow\bar{D}^{0}\pi^{0} l^{+}\nu_l \rightarrow K^{+} \pi^{-} \pi^{0} \pi^{0}  \pi^{0} l^{+}\nu_l $ & $\left(7.36\sim9.24\right)\times 10^{9}$   & $\left(3.94\sim4.94\right)\times 10^{8}$ \\
\hline
$B^{+} \rightarrow\bar{D}^{*}(2007)^0 l^{+}\nu_l \rightarrow\bar{D}^{0}\pi^{0} l^{+}\nu_l \rightarrow K^{+} \pi^{-} \pi^{-} \pi^{+} \pi^{0}  \pi^{0} l^{+}\nu_l $ & $\left(1.47\sim1.97\right)\times 10^{10}$   & $\left(0.79\sim1.05\right)\times 10^{9}$ \\
\hline
$B^{+} \rightarrow\bar{D}^{*}(2007)^0 l^{+}\nu_l \rightarrow\bar{D}^{0}\gamma l^{+}\nu_l \rightarrow K^{+} \pi^{-} \pi^{0}\gamma l^{+}\nu_l  $ & $\left(0.82\sim1.05\right)\times 10^{10}$   & $\left(4.41\sim5.61\right)\times 10^{8}$ \\
\hline
$B^{+} \rightarrow\bar{D}^{*}(2007)^0 l^{+}\nu_l \rightarrow\bar{D}^{0}\gamma l^{+}\nu_l \rightarrow K^{+} \pi^{-} \pi^{-}\pi^{+}\gamma l^{+}\nu_l  $ & $\left(1.45\sim1.83\right)\times 10^{10}$   & $\left(7.77\sim9.76\right)\times 10^{8}$ \\
\hline
$B^{+} \rightarrow\bar{D}^{*}(2007)^0 l^{+}\nu_l \rightarrow\bar{D}^{0}\gamma l^{+}\nu_l \rightarrow K^{+} \pi^{-} \pi^{0}\pi^{0}\gamma l^{+}\nu_l  $ & $\left(1.35\sim1.70\right)\times 10^{10}$   & $\left(7.20\sim9.07\right)\times 10^{8}$ \\
\hline
\end{tabular}
\end{center}
\end{table}

From the numerical results of $(\epsilon_f N)_{CP}^{pm}$ and $(\epsilon_f N)_{CP}^{mp}$ in Table~\ref{tabnmneedcpasympm}, we can see that the numbers of the $B_c^{\pm}$ events-times-efficiency, which are needed to observe the CP asymmetries ${\mathcal A}_{CP}^{pm}$ and ${\mathcal A}_{CP}^{mp}$ at three standard deviations (3$\sigma$) level, are of the order of $10^8 \sim 10^{10}$, so they are possible to be marginally observed at the LHC experiment and the HL-LHC experiment.
\subsection{The numerical results of the asymmetry observables ${\mathcal A}_{CP}^{pp}$ and ${\mathcal A}_{CP}^{mm}$}
Now, we move on to calculate the numerical results of the asymmetry observables ${\mathcal A}_{CP}^{pp}$ and ${\mathcal A}_{CP}^{mm}$. By substituting the values of the parameters in Eqs.(\ref{Eq:valparameters}) and (\ref{Eq:ckmwolfenparval}) into Eqs.(\ref{Eq:cpasymmacppp3}) and (\ref{Eq:cpasymmacpmm3}), we have
\begin{align}
&{\mathcal A}_{CP}^{pp}=(6.29\pm0.04)\times 10^{-3},\label{Eq:numasyacppp}\\
&{\mathcal A}_{CP}^{mm}=(-3.24\pm0.39)\times 10^{-4}.\label{Eq:numasyacpmm}
\end{align}
From the above equations, we can see that the numerical value of ${\mathcal A}_{CP}^{pp}$ is larger than that of ${\mathcal A}_{CP}^{mm}$, the reasons are as follows: the final states of the neutral kaon are $(\pi^{+} e^{-} \bar{\nu}_{e})$ and $(\pi^{-} e^{+} \nu_{e})$ in the asymmetry observables ${\mathcal A}_{CP}^{pp}$ and ${\mathcal A}_{CP}^{mm}$, respectively, so ${\mathcal A}_{CP}^{pp}$ and ${\mathcal A}_{CP}^{mm}$ depend on different mixing parameters, which can be seen in Eqs.(\ref{Eq:cpasymmacppp3}) and (\ref{Eq:cpasymmacpmm3}). By our calculation, the numerator of ${\mathcal A}_{CP}^{pp}$ can be expressed as the sum of   $\left|p\right|^2-\left|q\right|^2$ and $\left|q\right|^2\cdot \left(1- r_g\right)$, while the numerator of ${\mathcal A}_{CP}^{mm}$ can be expressed as the difference of $\left|p\right|^2-\left|q\right|^2$ and $\left|p\right|^2\cdot \left(1- r_g\right)$, as shown in Eqs.(\ref{Eq:cpasymmacppp3}) and (\ref{Eq:cpasymmacpmm3}), so the numerical value of ${\mathcal A}_{CP}^{pp}$ is larger than that of ${\mathcal A}_{CP}^{mm}$.

Using the same method as in subsection~\ref{sec:numrescpasymmetry}, the numbers of $B_c^{\pm}$ events-times-efficiency needed to establish the asymmetry observables ${\mathcal A}_{CP}^{pp}$ and ${\mathcal A}_{CP}^{mm}$ to three standard deviations (3$\sigma$) can be calculated by using the following formulas
\begin{align}
(\epsilon_f N)_{CP}^{pp}=&\frac{9}{\left[{\mathcal B}(B_c^+ \rightarrow B^{+} \pi^{+} e^{-} \bar{\nu}_{e}\rightarrow f_{B^+} \pi^{+} e^{-} \bar{\nu}_{e})+{\mathcal B}(B_c^- \rightarrow B^{-} \pi^{+} e^{-} \bar{\nu}_{e}\rightarrow f_{B^-} \pi^{+} e^{-} \bar{\nu}_{e})\right]\cdot \left| {\mathcal A}_{CP}^{pp}\right|}\nonumber\\
\approx &\frac{9}{2\cdot {\mathcal B}(B_{c}^{\pm}\rightarrow  B^{\pm} \pi^{+} e^{-} \bar{\nu}_{e})\cdot {\mathcal B}(B^+\rightarrow f_{B^+})\cdot \left|{\mathcal A}_{CP}^{pp}\right|},\label{Eq:numneedcppp}
\end{align}
and
\begin{align}
(\epsilon_f N)_{CP}^{mm}=&\frac{9}{\left[{\mathcal B}(B_c^+ \rightarrow B^{+} \pi^{-} e^{+} \nu_{e}\rightarrow f_{B^+} \pi^{-} e^{+} \nu_{e})+{\mathcal B}(B_c^- \rightarrow B^{-} \pi^{-} e^{+} \nu_{e}\rightarrow f_{B^-} \pi^{-} e^{+} \nu_{e})\right]\cdot \left| {\mathcal A}_{CP}^{mm}\right|}\nonumber\\
\approx &\frac{9}{2\cdot {\mathcal B}(B_{c}^{\pm}\rightarrow  B^{\pm} \pi^{-} e^{+} \nu_{e})\cdot {\mathcal B}(B^+\rightarrow f_{B^+})\cdot \left|{\mathcal A}_{CP}^{mm}\right|}.\label{Eq:numneedcpmm}
\end{align}
By substituting the numerical results of the asymmetry observables ${\mathcal A}_{CP}^{pp}$ and ${\mathcal A}_{CP}^{mm}$ in Eqs.(\ref{Eq:numasyacppp}) and (\ref{Eq:numasyacpmm}), the numerical results of ${\mathcal B}(B_{c}^{\pm}\rightarrow  B^{\pm} \pi^{+} e^{-} \bar{\nu}_{e})\cdot {\mathcal B}(B^+\rightarrow f_{B^+})$ in Table~\ref{totbranchratio3} and that of ${\mathcal B}(B_{c}^{\pm}\rightarrow  B^{\pm} \pi^{-} e^{+} \nu_{e})\cdot {\mathcal B}(B^+\rightarrow f_{B^+})$ in Table~\ref{totbranchratio4} into Eqs.(\ref{Eq:numneedcppp}) and (\ref{Eq:numneedcpmm}), we can obtain the numerical results of $(\epsilon_f N)_{CP}^{pp}$ and $(\epsilon_f N)_{CP}^{mm}$, which are listed in Table~\ref{tabnmneedasyppandmm}. Obviously, the numbers of the $B_c^{\pm}$ events-times-efficiency, which are needed to observe the asymmetry observables ${\mathcal A}_{CP}^{pp}$ and ${\mathcal A}_{CP}^{mm}$ at three standard deviations (3$\sigma$) level, are of the order of $10^8 \sim 10^{10}$, so they are hopefully to be marginally observed at the LHC experiment and the HL-LHC experiment.
\begin{table}[t]
\begin{center}
\caption{\label{tabnmneedasyppandmm} \small The numerical results of $(\epsilon_f N)_{CP}^{pp}$ and $(\epsilon_f N)_{CP}^{mm}$.}
\vspace{0.1cm}
\doublerulesep 0.8pt \tabcolsep 0.18in
\scriptsize
\begin{tabular}{c|c|c}
\hline
the $B^{+} \rightarrow f_{B^+}  $ decays & $(\epsilon_f N)_{CP}^{pp}$ & $(\epsilon_f N)_{CP}^{mm}$\\
\hline
$B^{+} \rightarrow\bar{D}^0 l^{+}\nu_l \rightarrow K^{+} \pi^{-} \pi^{0} l^{+}\nu_l $ & $\left(3.93\sim4.93\right)\times 10^{8}$   & $\left(0.72\sim1.01\right)\times 10^{10}$ \\
\hline
$ B^{+} \rightarrow\bar{D}^0 l^{+}\nu_l \rightarrow K^{+} \pi^{-} \pi^{-}\pi^{+} l^{+}\nu_l $ & $\left(6.92\sim8.57\right)\times 10^{8}$   & $\left(1.27\sim1.76\right)\times 10^{10}$ \\
\hline
$ B^{+} \rightarrow\bar{D}^0 l^{+}\nu_l \rightarrow K^{+} \pi^{-} \pi^{0}\pi^{0} l^{+}\nu_l $ & $\left(6.41\sim7.96\right)\times 10^{8}$   & $\left(1.18\sim1.64\right)\times 10^{10}$ \\
\hline
$B^{+} \rightarrow\bar{D}^{*}(2007)^0 l^{+}\nu_l \rightarrow\bar{D}^{0}\pi^{0} l^{+}\nu_l \rightarrow K^{+} e^{-} \bar{\nu}_{e} \pi^{0} l^{+}\nu_l $ & $\left(0.99\sim1.23\right)\times 10^{9}$    & $\left(1.82\sim2.53\right)\times 10^{10}$ \\
\hline
$B^{+} \rightarrow\bar{D}^{*}(2007)^0 l^{+}\nu_l \rightarrow\bar{D}^{0}\pi^{0} l^{+}\nu_l \rightarrow K^{+} \mu^{-} \bar{\nu}_{\mu} \pi^{0} l^{+}\nu_l $ & $\left(1.03\sim1.28\right)\times 10^{9}$    & $\left(1.89\sim2.63\right)\times 10^{10}$ \\
\hline
$B^{+} \rightarrow\bar{D}^{*}(2007)^0 l^{+}\nu_l \rightarrow\bar{D}^{0}\pi^{0} l^{+}\nu_l \rightarrow K^{+} \pi^{-} \pi^{0} l^{+}\nu_l $ & $\left(0.89\sim1.11\right)\times 10^{9}$    & $\left(1.64\sim2.27\right)\times 10^{10}$ \\
\hline
$B^{+} \rightarrow\bar{D}^{*}(2007)^0 l^{+}\nu_l \rightarrow\bar{D}^{0}\pi^{0} l^{+}\nu_l \rightarrow K^{+} \pi^{-}  \pi^{0} \pi^{0} l^{+}\nu_l $ & $\left(2.42\sim3.06\right)\times 10^{8}$    & $\left(4.47\sim6.26\right)\times 10^{9}$ \\
\hline
$B^{+} \rightarrow\bar{D}^{*}(2007)^0 l^{+}\nu_l \rightarrow\bar{D}^{0}\pi^{0} l^{+}\nu_l \rightarrow K^{+} \pi^{-} \pi^{-} \pi^{+}  \pi^{0} l^{+}\nu_l $ & $\left(4.26\sim5.32\right)\times 10^{8}$    & $\left(0.79\sim1.09\right)\times 10^{10}$ \\
\hline
$B^{+} \rightarrow\bar{D}^{*}(2007)^0 l^{+}\nu_l \rightarrow\bar{D}^{0}\pi^{0} l^{+}\nu_l \rightarrow K^{+} \pi^{-} \pi^{0} \pi^{0}  \pi^{0} l^{+}\nu_l $ & $\left(3.95\sim4.95\right)\times 10^{8}$   & $\left(0.73\sim1.01\right)\times 10^{10}$ \\
\hline
$B^{+} \rightarrow\bar{D}^{*}(2007)^0 l^{+}\nu_l \rightarrow\bar{D}^{0}\pi^{0} l^{+}\nu_l \rightarrow K^{+} \pi^{-} \pi^{-} \pi^{+} \pi^{0}  \pi^{0} l^{+}\nu_l $ & $\left(0.79\sim1.06\right)\times 10^{9}$   & $\left(1.47\sim2.15\right)\times 10^{10}$ \\
\hline
$B^{+} \rightarrow\bar{D}^{*}(2007)^0 l^{+}\nu_l \rightarrow\bar{D}^{0}\gamma l^{+}\nu_l \rightarrow K^{+} \pi^{-} \pi^{0}\gamma l^{+}\nu_l  $ & $\left(4.42\sim5.62\right)\times 10^{8}$   & $\left(0.82\sim1.15\right)\times 10^{10}$ \\
\hline
$B^{+} \rightarrow\bar{D}^{*}(2007)^0 l^{+}\nu_l \rightarrow\bar{D}^{0}\gamma l^{+}\nu_l \rightarrow K^{+} \pi^{-} \pi^{-}\pi^{+}\gamma l^{+}\nu_l  $ & $\left(7.79\sim9.77\right)\times 10^{8}$   & $\left(1.44\sim2.00\right)\times 10^{10}$ \\
\hline
$B^{+} \rightarrow\bar{D}^{*}(2007)^0 l^{+}\nu_l \rightarrow\bar{D}^{0}\gamma l^{+}\nu_l \rightarrow K^{+} \pi^{-} \pi^{0}\pi^{0}\gamma l^{+}\nu_l  $ & $\left(7.22\sim9.09\right)\times 10^{8}$   & $\left(1.33\sim1.86\right)\times 10^{10}$ \\
\hline
\end{tabular}
\end{center}
\end{table}
\section{Conclusions}
\label{sec:conclusions}
In this work, we derive the branching ratios for the $B_{c}^{\pm}\rightarrow  B^{\pm} \pi^{\pm} e^{\mp} \nu_{e}$ and $B_{c}^{\pm}\rightarrow  B^{\pm} \pi^{\mp} e^{\pm} \nu_{e}$ decays. Combining the numerical results of the branching ratios for these decays with the branching ratios for the consequent decays of the $B^{\pm}$ mesons given by the Particle Data Group, we obtain the total branching ratios of the decay chains $B_{c}^{\pm}\rightarrow  B^{\pm} \pi^{\pm} e^{\mp} \nu_{e}\rightarrow f_{B^{\pm}} \pi^{\pm} e^{\mp} \nu_{e}$ and $B_{c}^{\pm}\rightarrow  B^{\pm} \pi^{\mp} e^{\pm} \nu_{e}\rightarrow f_{B^{\pm}} \pi^{\mp} e^{\pm} \nu_{e}$. We find that some of the $B_{c}^{\pm}\rightarrow  B^{\pm} \pi^{\pm} e^{\mp} \nu_{e}\rightarrow f_{B^{\pm}} \pi^{\pm} e^{\mp} \nu_{e}$ and $B_{c}^{\pm}\rightarrow  B^{\pm} \pi^{\mp} e^{\pm} \nu_{e}\rightarrow f_{B^{\pm}} \pi^{\mp} e^{\pm} \nu_{e}$ decay chains have large branching ratios, whose maximum value can reach $2.66\times 10^{-6}$. We estimate the numbers of the $B_c^{\pm}$ events-times-efficiency needed to observe the $B_{c}^{\pm}\rightarrow  B^{\pm} \pi^{\pm} e^{\mp} \nu_{e}\rightarrow f_{B^{\pm}} \pi^{\pm} e^{\mp} \nu_{e}$ and $B_{c}^{\pm}\rightarrow  B^{\pm} \pi^{\mp} e^{\pm} \nu_{e}\rightarrow f_{B^{\pm}} \pi^{\mp} e^{\pm} \nu_{e}$ decays at three standard deviations ($3\sigma$) level and find that they are of the order of $10^6 \sim 10^{7}$, so these decays may be measurable at the LHC experiment.

We investigate the CP asymmetry observables ${\mathcal A}_{CP}^{pm}$ and ${\mathcal A}_{CP}^{mp}$ in $B_{c}^{\pm}\rightarrow  B^{\pm} \pi^{\pm} e^{\mp} \nu_{e}$ and $B_{c}^{\pm}\rightarrow  B^{\pm} \pi^{\mp} e^{\pm} \nu_{e}$ decays, which consist of three parts: the indirect CP violations in $K^0 -\bar{K}^0$ mixing $A_{CP,mix}^{pm}$ and $A_{CP,mix}^{mp}$, the direct CP violations in $B_c$ decay $A_{CP,dir}^{pm}$ and $A_{CP,dir}^{mp}$, the new CP violation effects $A_{CP,int}^{pm}$ and $A_{CP,int}^{mp}$, which originates from the interference between the amplitude of the $B_{c}^{-}\rightarrow B^{-} K^{0} (\bar{K}^0)\rightarrow B^{-} \pi^{-}  e^{+} \nu_{e} (\pi^{+}  e^{-} \bar{\nu}_{e})$ decay with the difference between the mixing effect of $K^{0}\rightarrow \bar{K}^0$ and that of $\bar{K}^0\rightarrow K^{0}$. The strong phase differences of the direct CP asymmetries $A_{CP,dir}^{pm}$ and $A_{CP,dir}^{mp}$ arising from $K^0-\bar{K}^0$ mixing parameters, and thus preventing pollution from strong dynamics.

We present the numerical results of the CP asymmetry observables ${\mathcal A}_{CP}^{pm}$ and ${\mathcal A}_{CP}^{mp}$ in $B_{c}^{\pm}\rightarrow  B^{\pm} \pi^{\pm} e^{\mp} \nu_{e}$ and $B_{c}^{\pm}\rightarrow  B^{\pm} \pi^{\mp} e^{\pm} \nu_{e}$ decays, we find that ${\mathcal A}_{CP}^{pm}$ and ${\mathcal A}_{CP}^{mp}$ are of the order $\mathcal{O}(10^{-4})$ and $\mathcal{O}(10^{-3})$, respectively. Meanwhile, the direct CP asymmetries ${\mathcal A}_{CP,dir}^{pm}$ and ${\mathcal A}_{CP,dir}^{mp}$ have a very small value and can be neglected safely. For the CP asymmetry observable ${\mathcal A}_{CP}^{pm}$ in $B_{c}^{\pm}\rightarrow  B^{\pm} \pi^{\pm} e^{\mp} \nu_{e}$ decays, the new CP violation effect ${\mathcal A}_{CP,int}^{pm}$ is dominant, the indirect CP violation in kaon mixing ${\mathcal A}_{CP,mix}^{pm}$ only accounts for about $5\%$ of the CP asymmetry observable ${\mathcal A}_{CP}^{pm}$. However, for the CP asymmetry observable ${\mathcal A}_{CP}^{mp}$ in $B_{c}^{\pm}\rightarrow  B^{\pm} \pi^{\mp} e^{\pm} \nu_{e}$ decays, the indirect CP violation in kaon mixing ${\mathcal A}_{CP,mix}^{mp}$ is dominant, the new CP violation effect ${\mathcal A}_{CP,int}^{mp}$ only accounts for about $5\%$ of the CP asymmetry observable ${\mathcal A}_{CP}^{mp}$. So the CP asymmetry observable ${\mathcal A}_{CP}^{pm}$ in $B_{c}^{\pm}\rightarrow  B^{\pm} \pi^{\pm} e^{\mp} \nu_{e}$ decays provides an ideal place to study the new CP violation effect.

Beside the CP asymmetry observables ${\mathcal A}_{CP}^{pm}$ and ${\mathcal A}_{CP}^{mp}$ in $B_{c}^{\pm} \rightarrow B^{\pm} \pi^{\pm} e^{\mp} \nu_{e}$ and $B_{c}^{\pm} \rightarrow B^{\pm} \pi^{\mp} e^{\pm} \nu_{e}$ decays, we also define another two asymmetry observables ${\mathcal A}_{CP}^{pp}$ and ${\mathcal A}_{CP}^{mm}$ in $B_{c}^+ \rightarrow B^+ \pi^{\pm} e^{\mp} \nu_{e}$ decays and their partially conjugated decays $B_{c}^- \rightarrow B^- \pi^{\pm} e^{\mp} \nu_{e}$. we derive the expressions for the asymmetry observables ${\mathcal A}_{CP}^{pp}$ and ${\mathcal A}_{CP}^{mm}$, which are dominated by $K^0-\bar{K}^0$ mixing. We calculate the numerical results of the asymmetry observables ${\mathcal A}_{CP}^{pp}$ and ${\mathcal A}_{CP}^{mm}$ and find that ${\mathcal A}_{CP}^{pp}$ is of the order $\mathcal{O}(10^{-3})$, while ${\mathcal A}_{CP}^{mm}$ is of the order $\mathcal{O}(10^{-4})$.

Together with the branching ratios for the $B_{c}^{\pm}\rightarrow  B^{\pm} \pi^{\pm} e^{\mp} \nu_{e}\rightarrow f_{B^{\pm}} \pi^{\pm} e^{\mp} \nu_{e}$, $B_{c}^{\pm}\rightarrow  B^{\pm} \pi^{\mp} e^{\pm} \nu_{e}\rightarrow f_{B^{\pm}} \pi^{\mp} e^{\pm} \nu_{e}$, $B_{c}^{\pm}\rightarrow  B^{\pm} \pi^{+} e^{-} \bar{\nu}_{e}\rightarrow f_{B^{\pm}} \pi^{+} e^{-} \bar{\nu}_{e}$ and $B_{c}^{\pm}\rightarrow  B^{\pm} \pi^{-} e^{+} \nu_{e}\rightarrow f_{B^{\pm}} \pi^{-} e^{+} \nu_{e}$ decay chains and the observables ${\mathcal A}_{CP}^{pm}$, ${\mathcal A}_{CP}^{mp}$, ${\mathcal A}_{CP}^{pp}$ and ${\mathcal A}_{CP}^{mm}$, we calculate the numbers of the $B_c^{\pm}$ events-times-efficiency needed to establish the observables ${\mathcal A}_{CP}^{pm}$, ${\mathcal A}_{CP}^{mp}$, ${\mathcal A}_{CP}^{pp}$ and ${\mathcal A}_{CP}^{mm}$ to three standard deviations in the decay processes with a large branching ratios. The numbers of the $B_c^{\pm}$ events-times-efficiency needed to observe the asymmetry observables ${\mathcal A}_{CP}^{pm}$, ${\mathcal A}_{CP}^{mp}$, ${\mathcal A}_{CP}^{pp}$ and ${\mathcal A}_{CP}^{mm}$ at a significance of 3$\sigma$ in the above mentioned decays are of the order of $10^8 \sim 10^{10}$, so these observables are hopefully to be marginally observed at the LHC experiment and the HL-LHC experiment.
\section*{Acknowledgements}
The work was supported by the National Natural Science Foundation of China (Contract Nos. 12175088).
\begin{appendix}
\numberwithin{equation}{section}
\section{The relations used in ${\mathcal A}_{CP}^{pm}$ and ${\mathcal A}_{CP}^{mp}$  calculations }
\label{sec:appendixcpcal}
Below we present the relations used in evaluating the expression of ${\mathcal A}_{CP}^{pm}$ and ${\mathcal A}_{CP}^{mp}$ in subsections~\ref{sec:cpcalacppm} and~\ref{sec:cpcalacpmp}
\begin{align}
&\Gamma(B_c^+\rightarrow B^{+}\pi^{\pm} e^{\mp} \nu_{e})\cdot\Gamma(B^{+}\rightarrow f_{B})-\Gamma(B_c^-\rightarrow B^{-}\pi^{\mp}  e^{\pm} \nu_{e})\cdot\Gamma(B^{-}\rightarrow \bar{f}_{B})\nonumber\\
&=\Gamma(B_c^+\rightarrow B^{+}\pi^{\pm} e^{\mp} \nu_{e})\cdot\left(\Gamma(B^{+}\rightarrow f_{B})-\Gamma(B^{-}\rightarrow \bar{f}_{B})\right)\nonumber\\
&~~~~~~~~~~~~~~~~+\left(\Gamma(B_c^+\rightarrow B^{+} \pi^{\pm} e^{\mp} \nu_{e})-\Gamma(B_c^-\rightarrow B^{-} \pi^{\mp}  e^{\pm} \nu_{e})\right)\cdot \Gamma(B^{-}\rightarrow \bar{f}_{B}),\label{Eq:bianhuanrelation1}
\end{align}
and
\begin{align}
&\Gamma(B_c^+\rightarrow B^{+}\pi^{\pm} e^{\mp} \nu_{e})\cdot\Gamma(B^{+}\rightarrow f_{B})+\Gamma(B_c^-\rightarrow B^{-}\pi^{\mp}  e^{\pm} \nu_{e})\cdot\Gamma(B^{-}\rightarrow \bar{f}_{B})\nonumber\\
&=\Gamma(B_c^+\rightarrow B^{+} \pi^{\pm} e^{\mp} \nu_{e})\cdot\left(\Gamma(B^{+}\rightarrow f_{B})+\Gamma(B^{-}\rightarrow \bar{f}_{B})\right)\nonumber\\
&~~~~~~~~~~~~~~~~-\left(\Gamma(B_c^+\rightarrow B^{+} \pi^{\pm} e^{\mp} \nu_{e})-\Gamma(B_c^-\rightarrow B^{-} \pi^{\mp}  e^{\pm} \nu_{e})\right)\cdot \Gamma(B^{-}\rightarrow \bar{f}_{B})\nonumber\\
&=\left(\Gamma(B_c^+\rightarrow B^{+} \pi^{\pm} e^{\mp} \nu_{e})+\Gamma(B_c^-\rightarrow B^{-} \pi^{\mp}  e^{\pm} \nu_{e}\right)\cdot \Gamma(B^{-}\rightarrow \bar{f}_{B})\nonumber\\
&~~~~~~~~~~~~~~~~+\Gamma(B_c^+\rightarrow B^{+} \pi^{\pm} e^{\mp} \nu_{e})\cdot\left(\Gamma(B^{+}\rightarrow f_{B})-\Gamma(B^{-}\rightarrow \bar{f}_{B})\right).\label{Eq:bianhuanrelation2}
\end{align}
\section{The relations used in ${\mathcal A}_{CP}^{pp}$ and ${\mathcal A}_{CP}^{mm}$  calculations }
In subsection~\ref{sec:asycalappmm}, the following equations are employed to calculate the expression of ${\mathcal A}_{CP}^{pp}$ and ${\mathcal A}_{CP}^{mm}$
\begin{align}
&\Gamma(B_c^+\rightarrow B^{+}\pi^{\pm} e^{\mp} \nu_{e})\cdot\Gamma(B^{+}\rightarrow f_{B})-\Gamma(B_c^-\rightarrow B^{-}\pi^{\pm} e^{\mp} \nu_{e})\cdot\Gamma(B^{-}\rightarrow \bar{f}_{B})\nonumber\\
&=\Gamma(B_c^+\rightarrow B^{+}\pi^{\pm} e^{\mp} \nu_{e})\cdot\left(\Gamma(B^{+}\rightarrow f_{B})-\Gamma(B^{-}\rightarrow \bar{f}_{B})\right)\nonumber\\
&~~~~~~~~~~~~~~~~+\left(\Gamma(B_c^+\rightarrow B^{+} \pi^{\pm} e^{\mp} \nu_{e})-\Gamma(B_c^-\rightarrow B^{-} \pi^{\pm} e^{\mp} \nu_{e})\right)\cdot \Gamma(B^{-}\rightarrow \bar{f}_{B}),\label{Eq:bianhuanrelappmm1}
\end{align}
and
\begin{align}
&\Gamma(B_c^+\rightarrow B^{+}\pi^{\pm} e^{\mp} \nu_{e})\cdot\Gamma(B^{+}\rightarrow f_{B})+\Gamma(B_c^-\rightarrow B^{-}\pi^{\pm} e^{\mp} \nu_{e})\cdot\Gamma(B^{-}\rightarrow \bar{f}_{B})\nonumber\\
&=\Gamma(B_c^+\rightarrow B^{+} \pi^{\pm} e^{\mp} \nu_{e})\cdot\left(\Gamma(B^{+}\rightarrow f_{B})+\Gamma(B^{-}\rightarrow \bar{f}_{B})\right)\nonumber\\
&~~~~~~~~~~~~~~~~-\left(\Gamma(B_c^+\rightarrow B^{+} \pi^{\pm} e^{\mp} \nu_{e})-\Gamma(B_c^-\rightarrow B^{-} \pi^{\pm} e^{\mp} \nu_{e})\right)\cdot \Gamma(B^{-}\rightarrow \bar{f}_{B})\nonumber\\
&=\left(\Gamma(B_c^+\rightarrow B^{+} \pi^{\pm} e^{\mp} \nu_{e})+\Gamma(B_c^-\rightarrow B^{-} \pi^{\pm} e^{\mp} \nu_{e}\right)\cdot \Gamma(B^{-}\rightarrow \bar{f}_{B})\nonumber\\
&~~~~~~~~~~~~~~~~+\Gamma(B_c^+\rightarrow B^{+} \pi^{\pm} e^{\mp} \nu_{e})\cdot\left(\Gamma(B^{+}\rightarrow f_{B})-\Gamma(B^{-}\rightarrow \bar{f}_{B})\right).\label{Eq:bianhuanrelappmm2}
\end{align}
\end{appendix}

\end{document}